\documentclass[useAMS,usenatbib,onecolumn]{mn2e}
\usepackage[]{graphicx}

\title{The spectral-temporal properties of the prompt pulses and rapid
decay phase of GRBs}
\author[R. Willingale et al.]
{R. Willingale$^1$\thanks{E-mail: rw@star.le.ac.uk}, F. Genet$^{2,3}$, J. Granot$^2$ and
P.T. O'Brien$^1$\\
$^1$Department of Physics and Astronomy, University of Leicester, LE1 7RH, UK\\
$^2$Center for Astrophysics Research, University of Hertfordshire, UK\\
$^3$Racah Institute of Physics, Hebrew University of Jerusalem, Israel.}
\begin{document}
\date{Accepted  Received ; in original form }

\pagerange{\pageref{firstpage}--\pageref{lastpage}} \pubyear{2009}

\maketitle
\label{firstpage}
\begin{abstract}
The prompt emission from GRBs is the brightest
electromagnetic emission known yet it's origin is not understood.
The flux density of individual prompt pulses of a GRB can be represented
by an analytical expression derived assuming the emission is from a thin,
ultra-relativistically expanding, uniform, spherical shell over a finite range
of radii. We present the results of fitting this analytical expression
to the lightcurves from the four standard {\em Swift} BAT energy bands and
two standard {\em Swift} XRT energy bands of 12 bursts.
The expression includes the High Latitude Emission (HLE) component
and the fits provide a rigourous demonstration that the HLE can explain
the Rapid Decay Phase (RDP) of the prompt emission.
The model also accommodates some aspects of energy-dependent lag and
energy-dependent pulse width, but there are features
in the data which are not well represented. Some pulses have
a hard, narrow peak which is not well fitted or a rise and
decay which is faster than expected using the standard indices
derived assuming synchrotron emission from internal shocks, although it might be
possible to accommodate these features using
a different emission mechanism within the same overall framework.
The luminosity of pulses is correlated with the peak energy
of the pulse spectrum, $L_{f}\propto (E_{peak}(1+z))^{1.8}$,
and anti-correlated with the time since ejection of the pulse,
$L_{f}\propto(T_{f}/(1+z))^{-2.0}$.
\end{abstract}


\section{Introduction}

The ability of the {\em Swift} satellite \citep{2004ApJ...611.1005G} to rapidly and
autonomously slew when the Burst Alert Telescope 
(BAT energy range 15-350 keV \citet{2005SSRv..120..143B}) detects a GRB
enables it to point the X-Ray Telescope
(XRT energy range 0.3-10 keV \citet{2005SSRv..120..165B}) at the target
within $\approx100$ seconds of the trigger. For many bursts the combination of the
BAT and XRT provides continuous broadband coverage of the prompt emission through to
the onset of the afterglow. A rapid decay phase (RDP)
immediately following the prompt emission 
is observed for the majority of bursts \citep{2006ApJ...642..389N} and this decay
appears as a smooth continuation of the prompt, both temporally and spectrally
\citep{2006ApJ...647.1213O}. This strongly suggests that the decay is the tail
of the prompt emission and several models have been proposed to account for the decay
(see \citet{2007ApJ...666.1002Z} and references therein)
but the most popular is so-called High Latitude
Emission (HLE) or ``naked'' GRB emission \citep{2000ApJ...541L..51K}.
If the prompt emission from an expanding
quasi-spherical shell suddenly turns off then radiation from increasingly larger
angles relative to the line of sight continue to reach the observer because of the added
path length (time delay) introduced by the curvature of the shell.
The spectrum seen at later (delayed) times is subject to a smaller
Doppler shift and the flux decays rapidly as the radiation is beamed away
from our line of sight, leading to a simple predicted
relation between the temporal decay index and the spectral index,
$\alpha=2+\beta$ where $F_{\nu}(t)\propto t^{-\alpha}\nu^{-\beta}$, that should hold
at late times when $t-t_{0}\gg\Delta t$, where $t_{0}$ and $\Delta t$ are the start time
and width of the pulse respectively.

Several authors have attempted to demonstrate that the RDP is consistent
with the expectations of HLE \citep{2007ApJ...663..407B,2007ApJ...666.1002Z}
but simplifying assumptions about $t_{0}$ and $\Delta t$ have to be made to
allow comparison of the model predictions and the observations.
Usually a parameterized form based on the late time asymptotic
behaviour of the HLE from a single spike or pulse in the prompt lightcurve
is assumed although it is used near the peak of the pulse before
the asymptotic form is appropriate and the zero time for the
powerlaw decay is chosen in an arbitary way.
Furthermore, most bursts consist of
several prompt pulses which somehow combine to produce the single HLE tail.
About 25\% of decays have no significant spectral evolution and 
\citet{2007ApJ...666.1002Z} 
suggest these are dominated by a simple curvature effect
while the rest show a clear hard-to-soft change as the flux decays which they 
modelled using an evolving exponential spectrum in which the cut-off energy 
$E_{c}=E_{0} (t/t_{0}-1)^{-\alpha_{2}}$.
What is required is a theoretical
model that incorporates the pulse profile, the evolution of the pulse spectrum,
the late HLE tail and summation of the emission from several pulses
so that the smooth continuation of the flux from the prompt phase in different 
energy bands can be predicted and compared with the BAT and XRT energy-resolved lightcurves.
Such a model could be used to fit both prompt pulses seen by the BAT
and later soft flares seen by the XRT in a consistent way.
Our primary aim was to test whether or not the HLE component incorporated into
a physically realistic model could explain the RDP in a rigourous way but
in addition we also confront other aspects of the model with the data. 
We have used the analytical expression
for the temporal-spectral profile of individual prompt pulses
from \citet{2009MNRAS.399.1328G} 
which is largely based on \citet{1998ApJ...494L..49S},
\citet{2005ApJ...631.1022G} and \citet{2008ApJ...677...92G}.
Full details about the model including a comparison with past attempts
to model the RDP with HLE and the properties of the model pulses
are given in \citet{2009MNRAS.399.1328G}.

Here we present a first attempt at fitting this model to the {\em Swift}
BAT and XRT lightcurves of
GRBs. We have selected 12 objects using the following criteria:
early coverage from the XRT including a good measurement of the RDP;
relatively simple BAT lightcurves; overlapping or near contiguous
data from the BAT and XRT giving good coverage of the transition from
the prompt emission into the RDP; an absence of large X-ray flares
which may obscure the RDP.

\section{The model function}

\begin{figure}
\begin{center}
\includegraphics[height=7cm,angle=-90]{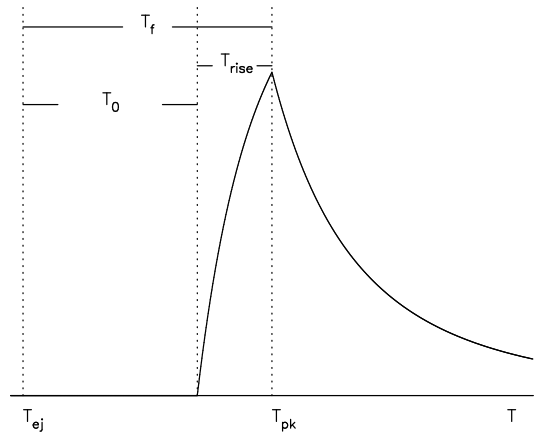}
\end{center}
\caption{The pulse profile showing the times $T_{ej}$ and $T_{pk}$ and the
timescales $T_{f}$ and $T_{0}=T_{f}-T_{rise}$.}
\label{fig1}
\end{figure}

Each individual peak or pulse in the prompt light curve is
assumed to originate from an
expanding thin shell ejected from the central engine modelled using
the emission profile given by \citet{2009MNRAS.399.1328G}.
For the case where the Lorentz factor of the expansion, $\Gamma$, is constant
with radius (coasting - index $m=0$ where $\Gamma^{2}\propto R^{-m}$)
the luminosity from within the shell in the comoving frame (indicated by
the primed variables) is $L'_{\nu'}=L'_{0}(R/R_{0})^{a}S(\nu'/\nu'_{p})$.
$S(x)$ is assumed to be the Band function
\citep{1993ApJ...413..281B}
and the peak frequency of the $\nu F_{\nu}$ spectrum is
$\nu'_{p}=\nu'_{0}(R/R_{0})^{d}$.
Under the standard internal shock model the index $d$ is linked with the
fact that the
emission process is synchrotron. Assuming the fast cooling regime
the peak frequency of the
$\nu F_{\nu}$ spectrum is $\nu'_m \propto B'\gamma_m^2$ where $B'$ is the
comoving
magnetic field and $\gamma_m$ is the electron energy (Lorentz factor)
at the peak of the spectrum.
As we are in the
coasting phase the strength of shocks is roughly constant, so $\gamma_m$ is
constant, and then $\nu'_p = \nu'_m \propto B'$.
$B'$ is predominantly normal to
the radial direction, so that $B' \approx B/\Gamma \propto B$ for $m=0$.
We expect $B\propto R^{-1}$, hence $\nu'_p \propto R^{-1}$, and thus $d=-1$. 
More generally, for uniform shells with a roughly
constant strength of the shocks, both the rate of particles crossing the
shock and the average energy per particle are constant with radius, implying
a constant rate of internal energy generation $dE'_{int}/dt'\propto R^0$. For
fast cooling this also applies to the total comoving luminosity
$L' \sim \nu'_p L'_{\nu'_p} \propto R^0$, so when $L'_{\nu'_p}$
evolves with radius so must
$\nu'_p$ and we must have $d+a=0$. In particular, this is true in the case of
synchrotron emission where $d=-1$ and $a=1$.
The emission turns on when the shell is at radius $R_{0}$ and turns
off at radius $R_{f}=R_{0}+\Delta R$.
If $m=0$ and $d=-1$ we can integrate the comoving luminosity over the
equal arrival time surface (EATS) giving
the number of photons $N$ per unit photon energy $E$, area $A$ at
observed time $T$ as
\begin{equation}
\frac{dN}{dEdAdT} (E,T\ge T_{ej}+T_{0})= P(T-T_{ej},T_{f},T_{rise})
B\left( \frac{E}{E_{f}} \frac{T-T_{ej}}{T_{f}} \right) 
\label{eq1}
\end{equation}
where the pulse profile is
\begin{equation}
P(T-T_{ej},T_{f},T_{rise})=
\left[ \left(min\left(\frac{T-T_{ej}}{T_{f}},1\right)^{a+2}
-\left(\frac{T_{f}-T_{rise}}{T_{f}}\right)^{a+2}\right)
\left(1-\left(\frac{T_{f}-T_{rise}}{T_{f}}\right)^{a+2}\right)^{-1}\right]
\left( \frac{T-T_{ej}}{T_{f}} \right) ^{-1}
\label{eq2}
\end{equation}
The characteristic times of the pulse $T_{f}$ and
$T_{0}=T_{f}-T_{rise}$ are the 
arrival times in the observer frame of the last photon and the
first photon emitted from
the shell, along the line of sight, measured
with respect to the the ejection time, $T_{ej}$.
The term in square brackets in Equation \ref{eq2} models the rise in the
pulse controlled by the temporal index $a$ and the timescales $T_{rise}$
and $T_{f}$. The functional
form of this rise is the same as in the original formulation
but we have included a normalisation such that the value
is 1 at $T-T_{ej}=T_{f}$ and have introduced the rise time
of the pulse $T_{rise}$.
Figure \ref{fig1} is a schematic
of the pulse profile showing the times and timescales.

$B(z)$ is the Band function
with $z=(E/E_{f})((T-T_{ej})/T_{f})^{-1}$, such that
\begin{equation}
B(z)=B_{norm}\left\{
\begin{array}{ll}
z^{b_{1}-1} e^{-z} & z \le b_{1}-b_{2}  \\
z^{b_{2}-1}(b_{1}-b_{2})^{b_{1}-b_{2}} e^{-(b_{1}-b_{2})} & z > b_{1}-b_{2}  \\
\end{array}
\right.
\end{equation}
The parameters of the Band function are the normalisation $B_{norm}$,
the low energy spectral index $b_{1}$, the characteristic energy $E_{c}$
and the high energy spectral index $b_{2}$.
The characteristic energy evolves with time,
$E_{c}(t)=E_{c}(T-T_{ej})=E_{f}((T-T_{ej})/T_{f})^{-1}$,
where $E_{f}$ is the value
of the cut-off energy at time $T_{pk}=T_{ej}+T_{f}$.
The peak of the $\nu F(\nu)$ spectrum at this time is
$E_{peak}=(b_{1}+1)E_{f}$.
$T_{pk}$ marks the start of the RDP and is usually the
peak of the pulse as indicated in Fig. \ref{fig1}
although if $T_{rise}$ is large, the index $d<-1$ and/or the spectrum
is hard the pulse maximum can occur before $T_{pk}$. 
So the index $d$ in the present formulation plays a similar role to
the index $\alpha_{2}$ in \citet{2007ApJ...666.1002Z} although we note
that the formulation presented here is rigorously valid only for $d=-1$,
since only in this case does the observed spectrum have the
same pure Band function form as the comoving emission spectrum.

The combination of the pulse profile function $P(t,T_{f},T_{rise})$ and
the time dependent blue-shift of the spectral profile $B(z)$ governed by
the index $d$ produce the rise
and fall of the pulse. If $d=-1$ the temporal decay index 
of the photon count rate after $t=T-T_{ej}=T_{f}$ 
at frequency $\nu$ is $\alpha_{\nu}=2+\beta_{\nu}$
where $\beta_{\nu}$ is the spectral index of the Band function at that
frequency. Thus
this formulation embodies the well known closure relation between the temporal
and spectral indices from the high latitude emission.
In the original formulation of \citet{2009MNRAS.399.1328G} the datum for the 
characteristic energy of the Band function was $E_{0}$ at time $T_{ej}+T_{0}$.
However, we have chosen to use $E_{f}$ at time $T_{pk}$ because
this time marks the maximum of the emission when we get the best 
measurement and estimate of the spectral profile.

Previous authors have attempted to model the prompt pulses of GRBs, e.g.
\citet{2005ApJ...627..324N}. The empirical pulse profile employed
incorporated an
exponential rise followed by an exponential decay and the profile was fitted
independently to each energy band to map any changes of the time constants
as a function of energy. The pulse profile used for the current work
is physically motivated and based on a simple but reasonably 
realistic model. The rise and fall of the pulse
are modelled by power laws and the spectral behaviour/evolution
is incorporated into the Doppler shift of the spectral profile as a function
of time.
The resulting temporal-spectral profile is fitted simultaneously to all
energy bands.

The prompt emission of a typical burst comprises several pulses and 
although the spectral-temporal evolution of each pulse in the model
is governed by the same simple formulation the evolution of a linear
superposition of pulses
can be complex. Each pulse can have a different spectrum
($E_{f}$, $b_{1}$ and $b_{2}$), different time scales ($T_{rise}$ and
$T_{f}$ and different brighness ($B_{norm}$).
The current model imposes no restriction
on the order or magnitude of the time scales or spectral parameters
in the pulse sequence and therefore the characteristic energy
of the combined spectrum can increase or decrease with time in
a complicated fashion.

\section{Model fitting}

We have attempted to fit the complete {\em Swift} BAT and
XRT light curves of the selected sample of 12 GRBs
using the above model for the prompt pulses.
An afterglow component fit to the plateau and final decay of the XRT lightcurve
as specified by \citet{2007ApJ...662.1093W}
was also included so that the transition from the initial decay
from the prompt into the plateau phase of the afterglow was modelled in a consistent way.
The standard energy bands were used for the BAT (15-25 keV, 25-50 keV,
50-100 keV and 100-350 keV) and XRT (0.3-1.5 keV and 1.5-10 keV).
XSPEC \citep{1996ASPC..101...17A} version 12.4.0 was used to calculate the
expected count rates in these energy bands from the prompt pulse
and afterglow components. The number of individual count rate predictions
required to find a best fit model is very large, $\sim 500000$ for
each pulse or afterglow component. Therefore using XSPEC directly
for each evaluation would have been prohibitively time consuming
(taking many hours to run on a desktop PC). 
Instead XSPEC was used to generate a look-up table of count rates
over a 3-D grid of spectral parameters for the Band function,
($b_{1}$, $b_{1}-b_{2}$ and $log_{10}(E_{c})$), and the fitting was
done by linear interpolation of the predicted count rates
within the look-up table. A look-up table
of $20\times20\times20$ samples typically takes 20 minutes to generate but
the fits can then be done in a few seconds. It was found that
the Band function model in XSPEC, Grbm, was unable to generate
spectra with $E_{c}<1.0$ keV because there is a hard limit set in the code.
Unfortunately late-time spectra from the pulse model above can easily
reach such low values of characteristic energy so
we had to create a local Band function model for XSPEC without this
restriction.

The count rates in the low energy XRT band are sensitive to
the assumed absorption column density. Fixed values of both
the Galactic and instrinsic column density were included in
the spectral model run in XSPEC using wabs (and zwabs if the redshift
was known). These column density values were
determined separately from spectral fits to the XRT data. For the
GRBs in the sample the early XRT data are dominated by prompt continuum which
may not be a simple power law and the intrinsic column density determined
using these data might be compromised. We used late time data when available
and for some bursts with larger intrinsic column density we tried running
the fits with different intrinsic column density values to see if this
influenced the results. The systematic errors introduced by uncertainties
in the column densities are smaller than the 90\% range
produced by the fitting so the fits remain essentially the same.

It was easy to make an initial guess for $T_{pk}$ and $T_{rise}$
for each pulse. Using this
guess a fit was then done letting $T_{pk}$, $T_{rise}$ and $T_{f}$ vary
to find a reasonable fit to the pulse profile and a best fit value for
$T_{pk}$.
This value of $T_{pk}$ was then fixed in all subsequent fitting to find
the best fit values of both the spectral and temporal parameters.
In some cases the rise of the pulse is not well constrained by the
data and for these bursts $T_{rise}$ was fixed.
There is in fact some redundancy between the $T_{rise}$ and $a$ parameters.
A fast rise can be achieved using $a=1$ and $T_{rise}$ small but can
also be fitted using a larger $T_{rise}$ and $a>1$. For most pulses
the number and quality of the data points over the rise is such that
it is not possible to fit these two parameters simultaneously.
For fast cooling we expect $a+d=0$ and in order to reduce the number of
free parameters we included this constraint in the fitting.
In most cases we fixed $a=1$ and hence $d=-1$.

The initial values assumed for $E_{f}$ and $b_{1}-b_{2}$ were 1000 keV
and 1 respectively.
This specifies a transition to the high energy power law 
at $E=(b_{1}-b_{2})E_{f}$=1000 keV.
The $b_{1}-b_{2}$ values derived for bursts observed with {\em BATSE}
have a considerable spread \citep{1993ApJ...413..281B} but 
$b_{1}-b_{2}=1$ is typical.
$E_{f}$ was allowed to float to a lower energy to find the best fit.
In a few cases the $\chi^{2}$ was very insensitive to $E_{f}$ and the
best fit energy exceeded 1000 keV. For these pulses $E_{f}$ was fixed at
1000 keV. There is some degeneracy/coupling between $E_{f}$ and the
spectral index $b_{1}$ and for most pulses the errors are too large
to provide independent estimates of these two parameters which
determine the spectral shape. In all cases we estimated the error
range for $b_{1}$ but we were only able to estimate an error range of
$E_{f}$ for a few pulses. In subsequent analysis we combine 
$E_{f}$ and $b_{1}$ to estimate the $E_{peak}$ value and use this
single parameter to characterise the hardness of the spectrum.
None of the fits is able to really
constrain $b_{1}-b_{2}$ because the BAT response doesn't extend to high
enough energies and the total count available from individual pulses is
usually considerably less than the total fluence. 
However, 3 soft pulses, pulse 2 in 050724 and both the pulses in 080805, gave
a significantly better fit using a cut-off powerlaw without a high
energy tail ($b_{1}-b_{2}\rightarrow\infty$).
For these pulses we investigated how low $b_{1}-b_{2}$ in the Band
function could be set before the fit was significantly worse than
the cut-off powerlaw. In these cases the fixed $b_{1}-b_{2}$ used
for the fit is effectively an estimate of the lower limit for the change in
index into the high energy tail of the Band function.

The fitting was performed using a search for minimum $\chi^{2}$ and the resulting minimum
$\chi^{2}$ found along with the number of degrees of freedom are listed in Table \ref{tab1}.
The product $F_{f}T_{f}^{2+b_{1}}$ for each pulse determines the relative
contribution it will make to the decaying tail of the prompt emission
and all the major pulses visible in the lightcurves were included in the
modelling.  Some faint pulses visible in the data were not included in the model
because they were too faint to contribute significantly to the RDP.
There are features in all the lightcurves which are not well fit by the model
and the
final $\chi^{2}_{\nu}$ values are not statistically acceptable.
Other ways in which the data don't conform
to the model can be seen in the plots of the fits provided below and will
be discussed later.

Some theoretical/physical constraints were ignored. 
Pulses were fitted independently with no attempt to introduce 
coupling between their parameters.
So it is possible that a slow pulse could start before a fast pulse
but peak after the same fast pulse. In the simple version of the
theoretical model with constant $\Gamma$ this is clearly not physically possible.
We return to this point in the discussion below.

\section{Fitting Results}

The Tables \ref{tab1} and \ref{tab2} list the fitted parameter values
obtained. The sequence of pulses used for each burst are numbered in
temporal order. When a parameter
was allowed to vary the 90\% confidence range is given in parentheses
with the best fit value in the centre.
In Table \ref{tab1},
$T_{pk}$ seconds is the time since the trigger of the start of
the final decay of the pulse at the elapsed time $T_{f}$ seconds after
ejection. This is usually the peak of the pulse but can be a little
later.
The last column provides the redshift values where available.
In Table \ref{tab2},
$b_{1}$ is the low spectral index of the Band function
(i.e. $b_{1}=\alpha+1$ where $\alpha$ is the low photon index in the original
formulation of the Band function) and 
$b_{1}-b_{2}$ is the difference between the low and high spectral
indices of the Band function.
$F_{f}$ is the flux in units of
$10^{-8}$ ergs cm$^{-2}$ s$^{-1}$ over the energy band
0.3 - 350 keV at time $T_{pk}$
and $E_{f}$ keV is the characteristic energy of the Band function at
that time. The peak energy of the $\nu F(\nu)$ spectrum at time
$T_{pk}$ is given by $E_{peak}=(b_{1}+1)E_{f}$ and of course this peak energy
evolves with time in the same way as the characteristic energy.
If $b_{1}<-1$ then there is no positive peak energy and
for these very soft pulses $E_{peak}$ is probably  below the XRT energy band.
For pulses where the 90\%
range of $b_{1}$ encompasses $-1$ we have not listed an $E_{peak}$ value.
Note, we quote the flux over the combined XRT and BAT energy band because
all the fits incorporate data from both instruments.
The last column of Table \ref{tab2}, $L_{f}$, is the isotropic luminosity
ergs s$^{-1}$
at time $T_{pk}$ over the bolometic energy band 1-10000 keV.
This was estimated from the flux $F_{f}$ using the redshift and assuming
a cosmology with $H_{0}=71$ km s$^{-1}$
Mpc$^{-1}$, $\Omega=0.27$ and $\Lambda=0.73$ to calculate the
luminosity distance $d_{L}$.

We want to stress that the fitted $E_{peak}$ values listed in Table \ref{tab2}
are the value of this spectral parameter at a specific time, $T_{pk}$
seconds after the BAT trigger.
We are able to obtain such values from a Band function fit
because we are fitting both
the spectral profile and the temporal profile of the pulse simultaneously.
During a typical pulse the peak energy will range from a factor of $\sim1.5$,
greater than this value, at the start of the rise, to a factor
$\sim 0.2$, smaller
than this value, when the tail of the pulse disappears into the noise or
is obscured by a subsequent pulse. The average pulse spectrum expected
from the model is a weighted sum or convolution of a series of Band profiles.
If the pulse spectrum summed over the main peak
were fitted with a Band function we would get
an average peak energy values, $\bar{E}_{peak}$, which
would be similar to but not the same as the $E_{peak}$ values derived by
the current technique.
Actually, most pulses are too weak to provide sufficient statistics to fit
a Band function to the average pulse spectrum seen in the {\em Swift} BAT.
However, \citet{2009ApJ...704.1405K} were able to fit spectra over relatively
small time intervals using a combination of {\em Swift} BAT and
{\em Suzaku} WAM data and \citet{2007ApJ...663.1125P} provide
a listing of $E_{peak}$ values derived for short time intervals of
061121 from cutoff powerlaw fits to {\em Konus-Wind} data.
In Table \ref{tab3} we list a small sub-set of
the peak energy results which are common between the present work
and these studies.
The $\bar{E}_{peak}$ are indeed similar to our $E_{peak}$ values
although they tend to be a little lower because the average pulse
spectrum is a little softer than that at the peak. When a soft
pulse is followed by a hard pulse as is the case for pulses 4 and 5
of 061121 then the $\bar{E}_{peak}$ value increases as expected.
The second time interval for 061121 contains pulse 4 ($E_{peak}=288$)
and the start of pulse 5 ($E_{peak}=849$)
so the $\bar{E}_{peak}=608$ is
a blend of the two. The third time interval for 061121 contains
just pulse 5 and is a little harder with $\bar{E}_{peak}=621$
but not as hard as the peak value for pulse 5.

The fits for all 12 bursts included are illustrated by the
Figures \ref{fig2} through \ref{fig13} in date order.
The BAT count rates are per detector.
The observed total count rate is much larger
than these values and depends on the position of the burst within the BAT
field of view. The sub-sections below
provide a brief commentary on all the fits including a description of
the features which are dominating the $\Delta\chi$ distribution in each case.
We have studied the residuals in detail to
see if it likely that the fitted parameters and the estimated
confidence ranges are compromised by the rather high reduced
Chi-squared values. 
In three cases the quoted error range may
be too small because of systematic problems with the fitting. These are
the 2nd pulse in 050724 in the 15-25 keV band,
the 1st pulse in 060211A in the  50-100 keV band
and the 1st pulse of 080805 
in the 15-25 keV band.
In these pulses the observed flux is 
systematically higher than the model in the quoted band over 
a significant fraction of the peak.

\begin{table}
 \begin{tabular}{llcccccc}
 GRB & pulse & $T_{pk}$ & $a$ & $T_{rise}$ & $T_{f}$ & $z$ & $\chi^{2}/ndof$ \\
 \hline
 050724 & 1 &   0.0 &  1.00 &   1.1& (  1.8,  2.1,  2.7)
  
 &  0.258 &  1433.9/ 235 \\
 & 2 &  93.4 &  1.00 &  35.0& ( 63.6, 66.5, 70.1)
  
 & & \\
 & 3 & 0.50E+05 &  1.00 & 0.18E+05& (0.30E+05,0.43E+05,0.58E+05)
  
 & & \\
 \hline
 050814 & 1 &   5.7 &  1.00& ( 10.5, 11.8, 14.1)& ( 26.4, 31.5, 37.5)
  
 &  5.300 &   299.1/ 197 \\
 & 2 &  66.4 &  1.00 &  40.0& ( 59.3, 64.3, 71.0)
  
 & & \\
 \hline
 051001 & 1 &  -8.7 &  1.00 &  15.2& ( 65.1, 76.3, 91.2)
  
 & &   578.9/ 233 \\
 & 2 & 145.7 &  1.00& ( 40.1, 42.8, 46.2)& ( 64.8, 69.3, 75.1)
  
 & & \\
 \hline
 060211A & 1 &  83.3 &  1.00& ( 36.2, 43.7, 47.6)& ( 66.9, 74.5, 80.8)
  
 & &   446.5/ 246 \\
 & 2 & 158.6 &  1.00& ( 19.2, 21.5, 24.9)& ( 31.6, 35.9, 40.7)
  
 & & \\
 \hline
 060814 & 1 &  10.1 &  1.00 &  13.0& ( 13.7, 14.3, 15.1)
  
 &  0.840 &  2709.8/1146 \\
 & 2 &  16.5 &  1.00& (  3.0,  3.1,  3.3)& (  9.1,  9.8, 10.4)
  
 & & \\
 & 3 &  69.7 &  1.00& (  8.6,  8.9,  9.1)& ( 25.1, 26.1, 27.2)
  
 & & \\
 & 4 & 128.7 &  1.00 &   7.7& ( 34.3, 38.2, 42.8)
  
 & & \\
 & 5 & 150.1 &  1.00& ( 67.0, 73.9, 89.0)& (232.6,252.5,279.1)
  
 & & \\
 \hline
 061110A & 1 &   8.9 &  1.00& ( 20.9, 21.9, 23.1)& ( 36.1, 39.2, 42.4)
  
 &  0.758 &   417.3/ 174 \\
 & 2 & 140.0 &  1.00& ( 71.8, 75.1, 78.9)& (154.9,165.5,176.3)
  
 & & \\
 \hline
 061121 & 1 &   2.9 &  1.00& (  2.8,  3.3,  4.2)& (  4.0,  4.6,  5.7)
  
 &  1.314 &  3052.4/ 630 \\
 & 2 &  61.9 &  1.00& (  0.9,  1.0,  1.1)& (  8.6,  9.4, 10.4)
  
 & & \\
 & 3 &  68.6 &  1.00& (  2.1,  2.1,  2.2)& (  5.9,  6.5,  7.0)
  
 & & \\
 & 4 &  73.3 &  1.00& (  2.2,  2.2,  2.3)& (  2.8,  3.0,  3.2)
  
 & & \\
 & 5 &  75.0& ( 2.10, 2.39, 2.82) &   1.3& (  1.5,  1.7,  2.0)
  
 & & \\
 & 6 & 110.0 &  1.00& (  4.5,  4.8,  5.6)& ( 19.5, 22.6, 27.3)
  
 & & \\
 \hline
 061222A & 1 &   1.0 &  1.00 &   5.4& ( 11.4, 15.1, 20.5)
  
 & &  2010.2/ 702 \\
 & 2 &  25.7 &  1.00 &   1.0& (  4.2,  5.1,  6.2)
  
 & & \\
 & 3 &  60.0 &  1.00& (  7.1,  7.5,  7.9)& ( 15.7, 17.3, 19.0)
  
 & & \\
 & 4 &  73.8 &  1.00& (  4.0,  4.5,  5.0)& ( 15.8, 17.0, 18.4)
  
 & & \\
 & 5 &  84.7 &  1.00& (  4.5,  5.0,  5.1)& (  5.1,  5.5,  6.2)
  
 & & \\
 & 6 &  87.1 &  1.00& (  1.0,  1.1,  1.2)& (  2.5,  2.9,  3.3)
  
 & & \\
 & 7 &  90.0 &  1.00 &   0.5& (  0.8,  1.2,  1.6)
  
 & & \\
 & 8 & 120.0 &  1.00& (  3.0,  5.7,  8.1)& ( 52.6, 65.8, 85.6)
  
 & & \\
 \hline
 070420 & 1 & -35.0 &  1.00 &   9.8& ( 10.6, 13.2, 16.6)
  
 & &   982.9/ 407 \\
 & 2 & -15.5 &  1.00 &  13.5& ( 14.9, 17.4, 20.0)
  
 & & \\
 & 3 &   2.4 &  1.00& (  7.1,  7.7,  8.3)& ( 13.6, 14.6, 15.6)
  
 & & \\
 & 4 &  60.9 &  1.00& ( 12.7, 13.2, 13.9)& ( 18.1, 19.3, 20.5)
  
 & & \\
 \hline
 070621 & 1 &   2.0 &  1.00 &   5.0& ( 10.4, 12.2, 14.3)
  
 & &   602.8/ 303 \\
 & 2 &   8.0 &  1.00 &   1.2& (  4.7,  7.8, 12.1)
  
 & & \\
 & 3 &  12.0 &  1.00 &   1.9& (  2.7,  4.3,  7.3)
  
 & & \\
 & 4 &  18.0 &  1.00 &   2.8& ( 19.9, 21.7, 23.8)
  
 & & \\
 & 5 &  22.0 &  1.00 &   0.8& (  3.3,  5.0,  7.3)
  
 & & \\
 & 6 & 120.8 &  1.00 &  28.3& ( 96.3,109.2,123.4)
  
 & & \\
 \hline
 080229A & 1 &   0.3 &  1.00 &   1.5& (  7.0,  8.8, 10.8)
  
 & &  1494.4/ 444 \\
 & 2 &  28.0 &  1.00& (  1.5,  1.7,  2.0)& (  5.2,  7.9, 10.6)
  
 & & \\
 & 3 &  35.0 &  1.00& (  2.7,  2.9,  3.0)& (  6.2,  7.2,  8.0)
  
 & & \\
 & 4 &  38.6 &  1.00& (  0.7,  0.7,  0.8)& (  4.7,  5.3,  6.1)
  
 & & \\
 & 5 &  40.7 &  1.00& (  0.1,  0.1,  0.1)& (  1.5,  2.0,  2.9)
  
 & & \\
 & 6 &  55.6 &  1.00 &   1.6& ( 10.1, 11.0, 12.7)
  
 & & \\
 & 7 & 120.5 &  1.00& ( 21.2, 22.8, 24.5)& ( 63.4, 78.0, 90.9)
  
 & & \\
 \hline
 080805 & 1 &   1.5 &  1.00& (  5.9,  6.5,  7.2)& ( 37.2, 42.5, 47.4)
  
 &  1.505 &   677.0/ 251 \\
 & 2 & 124.4 &  1.00 &  51.9& ( 57.4, 65.7, 75.3)
  
 & & \\
 \hline
 \end{tabular}

\caption{Temporal parameters for the fitted pulses. $T_{pk}$, $T_{rise}$
and $T_{f}$ are in seconds.
A single value indicates a parameter which
was fixed during fitting. The 90\% range and the best fit value are given for parameters which
were allowed to float in the fitting.
The redshifts were taken from
\citet{2005GCN..3700....1P}, \citet{2006GCN..5826....1B},\citet{2007GCN..6759....1F},\citet{2008GCN..8077....1J},\citet{2007GCN..6663....1T} and
\citet{2006AIPC..836..552J}.
}
\label{tab1}
\end{table}
\begin{table}
 \begin{tabular}{llcccccc}
 GRB & pulse & $F_{f}$ & $E_{f}$ & $b_{1}$ & $b_{1}-b_{2}$ & $E_{peak}$ & $L_{f}
 $
 \\
 \hline
 050724 & 1 & ( 65.5, 75.9,173.1) & 1000. & (-0.94,-0.71,-0.66)
  &  1.00 &  289.$\pm$ 142.
 & (2.13$\pm$1.51)$10^{50}$ \\
 & 2 & ( 1.72, 1.85, 2.02) & ( 2.18, 2.48, 2.74) & ( 1.02, 1.11, 1.26)
  &  3.00 &   5.2$\pm$  0.7
 & (3.72$\pm$0.30)$10^{48}$ \\
 & 3 & (0.00052,0.00068,0.00084) &  186. & (-1.00,-0.68,-0.34)
  &  1.00 &
 & (1.47$\pm$0.35)$10^{45}$ \\
 \hline
 050814 & 1 & (  6.6, 10.5, 23.8) & 1000. & (-1.03,-0.79,-0.54)
  &  1.00 &
 & (4.16$\pm$3.42)$10^{52}$ \\
 & 2 & ( 4.32, 5.44, 6.73) &  637. & (-1.25,-1.19,-1.13)
  &  1.00 &
 & (2.05$\pm$0.46)$10^{52}$ \\
 \hline
 051001 & 1 & ( 2.66, 3.19, 5.48) &  540. & (-0.96,-0.80,-0.60)
  &  1.00 &  108.$\pm$  98. \\
 & 2 & ( 4.73, 5.52, 6.34) & (  10.,  12.,  15.) & ( 0.15, 0.22, 0.30)
  &  1.00 &   15.$\pm$   3. \\
 \hline
 060211A & 1 & ( 4.03, 5.12, 6.72) & 1000. & (-1.12,-1.03,-0.96)
  &  1.00 & \\
 & 2 & ( 3.67, 4.80, 6.30) &  240. & (-0.73,-0.66,-0.58)
  &  1.00 &   82.$\pm$  19. \\
 \hline
 060814 & 1 & ( 49.8, 56.3, 63.9) & 1000. & (-0.45,-0.41,-0.38)
  &  1.00 &  589.$\pm$  34.
 & (4.01$\pm$0.51)$10^{51}$ \\
 & 2 & ( 79.0, 92.1,113.6) &  694. & (-0.48,-0.42,-0.35)
  &  1.00 &  403.$\pm$  43.
 & (5.70$\pm$1.07)$10^{51}$ \\
 & 3 & ( 38.1, 42.4, 47.5) & ( 146., 183., 235.) & (-0.41,-0.37,-0.32)
  &  1.00 &  116.$\pm$  29.
 & (2.00$\pm$0.22)$10^{51}$ \\
 & 4 & ( 6.86, 8.45, 9.97) & (   0., 167., 329.) & (-0.35,-0.30,-0.27)
  &  1.00 &  117.$\pm$ 116.
 & (4.19$\pm$0.77)$10^{50}$ \\
 & 5 & (0.525,0.570,0.615) &  120. & (-1.55,-1.43,-1.34)
  &  1.00 &
 & (1.46$\pm$0.12)$10^{49}$ \\
 \hline
 061110A & 1 & ( 6.14, 8.67,12.82) & 1000. & (-0.90,-0.79,-0.70)
  &  1.00 &  206.$\pm$ 100.
 & (2.90$\pm$1.12)$10^{50}$ \\
 & 2 & (0.494,0.517,0.541) &   87. & (-2.78,-2.64,-2.52)
  &  1.00 &
 & (4.82$\pm$0.22)$10^{48}$ \\
 \hline
 061121 & 1 & ( 22.9, 29.3, 48.2) &  685. & (-0.70,-0.63,-0.64)
  &  1.00 &  251.$\pm$  20.
 & (4.26$\pm$1.84)$10^{51}$ \\
 & 2 & (  94., 112., 133.) &  625. & (-0.44,-0.40,-0.35)
  &  1.00 &  377.$\pm$  27.
 & (2.08$\pm$0.36)$10^{52}$ \\
 & 3 & ( 130., 158., 195.) &  563. & (-0.37,-0.32,-0.27)
  &  1.00 &  381.$\pm$  30.
 & (3.10$\pm$0.63)$10^{52}$ \\
 & 4 & ( 158., 176., 203.) &  374. & (-0.27,-0.23,-0.20)
  &  1.00 &  288.$\pm$  13.
 & (3.39$\pm$0.44)$10^{52}$ \\
 & 5 & ( 438., 519., 605.) &  698. & ( 0.19, 0.22, 0.26)
  &  1.00 &  849.$\pm$  23.
 & (2.64$\pm$0.42)$10^{53}$ \\
 & 6 & ( 2.03, 2.33, 3.75) &   50. & (-1.93,-1.64,-1.40)
  &  1.00 &
 & (1.89$\pm$0.70)$10^{50}$ \\
 \hline
 061222A & 1 & ( 7.39, 9.36,43.08) &  854. & (-0.92,-0.53,-0.11)
  &  1.00 &  398.$\pm$ 344. \\
 & 2 & ( 23.5, 43.2, 81.8) & 1000. & (-0.79,-0.62,-0.45)
  &  1.00 &  380.$\pm$ 166. \\
 & 3 & ( 17.2, 18.7, 27.9) & 1000. & (-0.84,-0.72,-0.66)
  &  1.00 &  279.$\pm$  90. \\
 & 4 & ( 15.7, 21.7, 30.0) & 1000. & (-0.77,-0.68,-0.59)
  &  1.00 &  324.$\pm$  88. \\
 & 5 & ( 70.9, 75.2, 96.3) &  197. & ( 0.10, 0.17, 0.25)
  &  1.00 &  230.$\pm$  15. \\
 & 6 & ( 113., 165., 186.) &  573. & (-0.13,-0.03, 0.00)
  &  1.00 &  554.$\pm$  38. \\
 & 7 & ( 37.6, 44.2,110.0) &  466. & (-0.47,-0.18, 0.17)
  &  1.00 &  384.$\pm$ 149. \\
 & 8 & (0.484,0.564,0.698) & ( 1.09, 4.89, 8.42) & (-1.65,-1.63,-0.88)
  &  1.00 & \\
 \hline
 070420 & 1 & ( 15.9, 37.9, 60.2) &  900. & (-0.29,-0.07, 0.18)
  &  1.00 &  837.$\pm$ 211. \\
 & 2 & ( 39.0, 44.4, 65.4) & 1000. & (-0.36,-0.18,-0.02)
  &  1.00 &  822.$\pm$ 170. \\
 & 3 & ( 105., 133., 184.) & 1000. & (-0.37,-0.29,-0.20)
  &  1.00 &  710.$\pm$  84. \\
 & 4 & ( 61.4, 85.3,126.9) & 1000. & (-2.06,-1.92,-1.83)
  &  1.00 & \\
 \hline
 070621 & 1 & ( 40.1, 43.8, 77.3) &  765. & (-0.45,-0.31,-0.15)
  &  1.00 &  530.$\pm$ 115. \\
 & 2 & ( 19.9, 25.2, 41.1) & 1000. & (-1.08,-0.68,-0.68)
  &  1.00 &  316.$\pm$ 201. \\
 & 3 & ( 21.2, 28.4,136.4) & 1000. & (-1.25,-1.00,-0.85)
  &  1.00 & \\
 & 4 & ( 57.3, 70.1,117.3) & 1000. & (-1.51,-1.38,-1.20)
  &  1.00 & \\
 & 5 & ( 25.7, 31.6, 92.9) &  731. & (-0.53,-0.21,-0.15)
  &  1.00 &  578.$\pm$ 140. \\
 & 6 & (0.482,0.536,0.594) &  387. & (-1.78,-1.62,-1.47)
  &  1.00 & \\
 \hline
 080229A & 1 & ( 54.9, 61.2,123.5) & 1000. & (-0.58,-0.38,-0.17)
  &  1.00 &  615.$\pm$ 204. \\
 & 2 & ( 11.4, 70.0,244.3) &  660. & (-1.36,-0.96,-0.87)
  &  1.00 & \\
 & 3 & ( 115., 160., 230.) &  503. & (-0.91,-0.81,-0.74)
  &  1.00 &   97.$\pm$  44. \\
 & 4 & ( 327., 560., 911.) &  420. & (-1.51,-1.36,-1.20)
  &  1.00 & \\
 & 5 & (  44., 179.,1056.) &  517. & (-1.72,-0.98,-0.94)
  &  1.00 & \\
 & 6 & ( 143., 260., 715.) & 1000. & (-2.24,-2.05,-1.94)
  &  1.00 & \\
 & 7 & (0.653,0.761,0.979) & ( 0.80, 2.52, 5.08) & (-2.22,-1.96,-1.67)
  &  1.00 & \\
 \hline
 080805 & 1 & (  2.6, 15.1, 25.4) & (  48.,  60.,  86.) & ( 0.50, 0.80, 0.81)
  &  2.50 &  108.$\pm$  36.
 & (2.86$\pm$2.15)$10^{51}$ \\
 & 2 & (0.379,0.483,0.528) & ( 1.14, 1.48, 1.61) & ( 0.77, 1.05, 1.06)
  &  2.50 &   3.0$\pm$  0.5
 & (7.46$\pm$1.15)$10^{49}$ \\
 \hline
 \end{tabular}

\caption{Spectral parameters for the fitted pulses. $F_{f}$ 
$10^{-8}$ ergs cm$^{-2}$ s$^{-1}$ over the energy band 0.3-350 keV is the flux at peak,
$E_{f}$ keV is the cut-off energy at peak, $E_{peak}=(b_{1}+1)E_{f}$ keV
is the energy peak of the $\nu F(\nu)$ spectrum and
$L_{f}$ ergs s$^{-1}$ the isotropic peak luminosity.
A single value indicates a parameter which
was fixed during fitting. The 90\% range and the best fit value are given for parameters which
were allowed to float in the fitting.}
\label{tab2}
\end{table}
\begin{table}
 \begin{tabular}{llcccc}
 GRB & pulse & $T_{pk}$ s & $E_{peak}$ keV& Interval s &
 $\bar{E}_{peak}$ keV \\
 \hline
 060814 & 1 &  10.1 & $589\pm34$  &  -11.75-10.75 & $365^{+119}_{-95}$ \\
        & 2 &  16.5 & $403\pm43$  &  10.75-30.75  & $350^{+136}_{-100}$ \\
 \hline
 061121 & 2 &  61.9 & $377\pm27$  &  61.876-70.324 & $478^{+158}_{-99}$ \\
        & 3 &  68.6 & $381\pm30$  &  \\
        & 4 &  73.7 & $288\pm13$  &  70.324-75.158 & $608^{+87}_{-71}$ \\
        & 5 &  75.0 & $849\pm23$  &  75.188-83.380 & $621^{+282}_{-159}$ \\
 \hline
 061222A & 3 &  60.0 & $279\pm90$ &  34.29-89.29  & $415^{+210}_{-159}$ \\
         & 4 &  73.8 & $324\pm88$ &  \\ 
         & 5 &  83.0 & $230\pm15$ &  75.29-99.79  & $248^{+163}_{-68}$ \\
         & 6 &  87.0 & $554\pm38$ &  80.29-89.79  & $299^{+122}_{-80}$ \\
         & 7 &  90.0 & $384\pm149$ &  \\
 \hline
 \end{tabular}

\caption{Comparison of $E_{peak}$ keV at time $T_{pk}$
derived in the present work with
average values, $\bar{E}_{peak}$ keV, derived for a Band function
using combined {\em Swift}/BAT and {\em Suzaku} Wide-band All-Sky Monitor (WAM)
data \citep{2009ApJ...704.1405K} and cutoff powerlaw fits
to {\em Konus-Wind} data \citep{2007ApJ...663.1125P}. The errors and
error ranges quoted are 90\% confidence.
Both $T_{pk}$ and Interval are time since BAT trigger.}
\label{tab3}
\end{table}

\subsection{GRB050724}

\begin{figure}
\begin{center}
\includegraphics[height=15cm,angle=0]{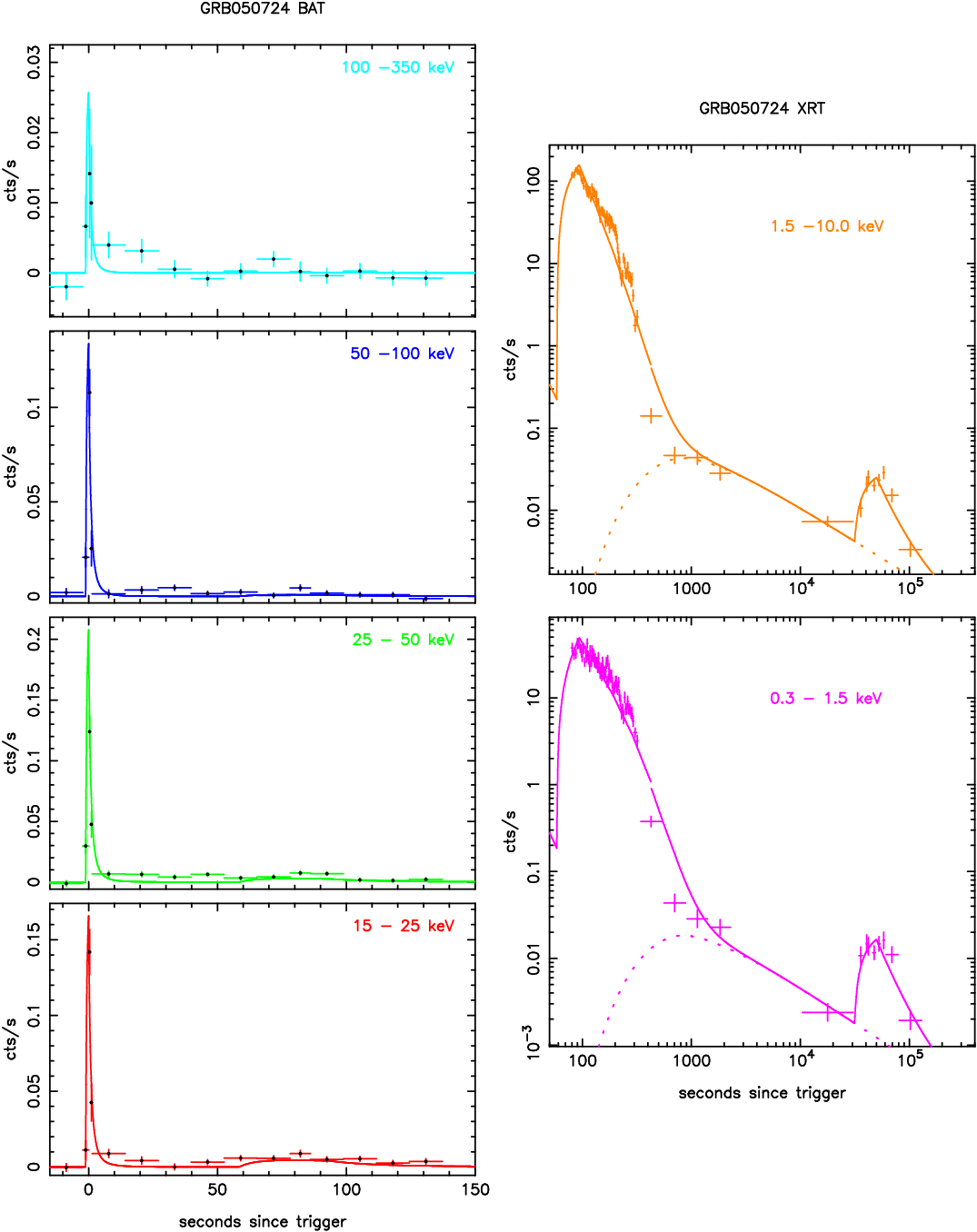}
\end{center}
\caption{GRB050724}
\label{fig2}
\end{figure}

\begin{figure}
\begin{center}
\includegraphics[height=15cm,angle=0]{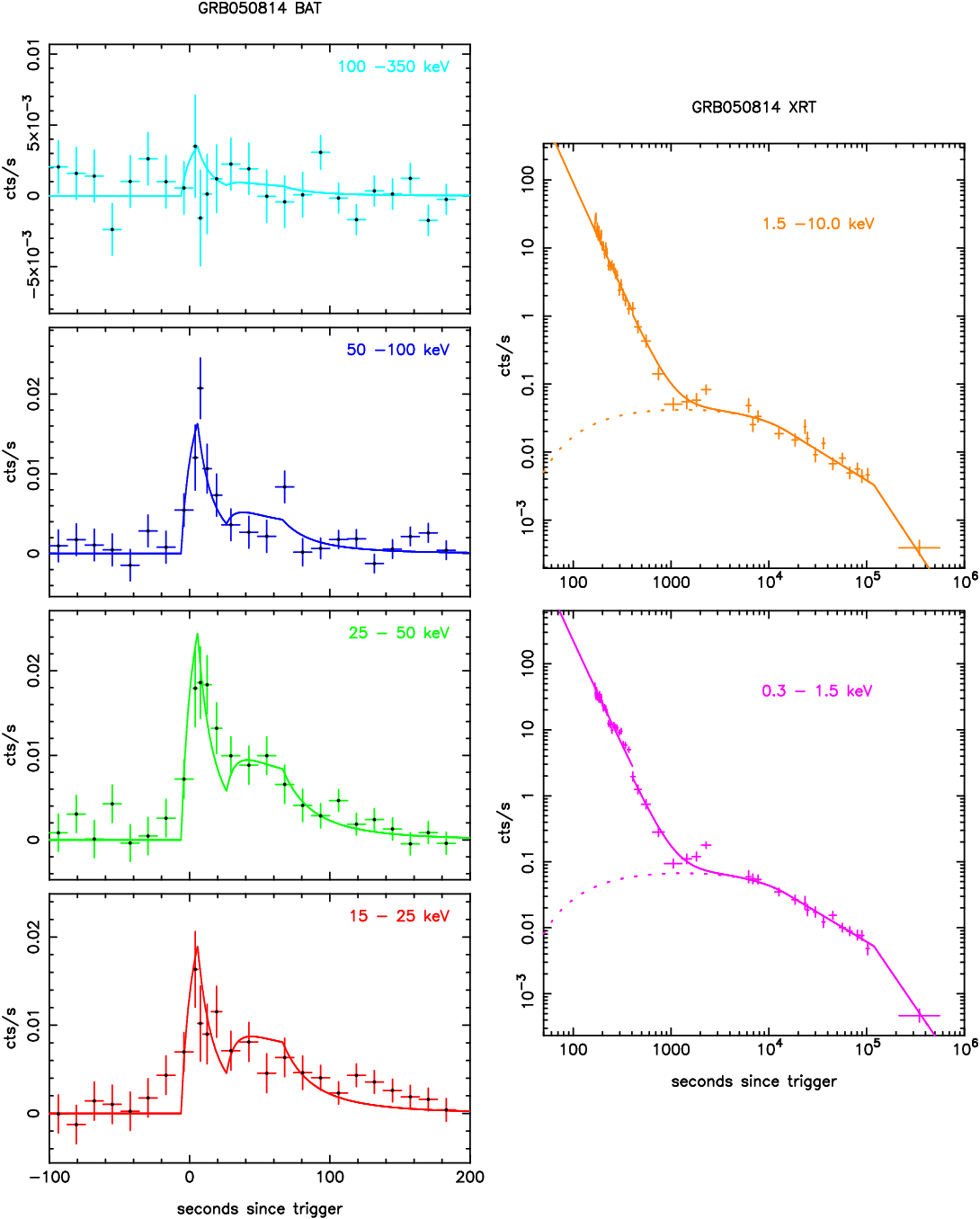}
\end{center}
\caption{GRB050814}
\label{fig3}
\end{figure}

\begin{figure}
\begin{center}
\includegraphics[height=15cm,angle=0]{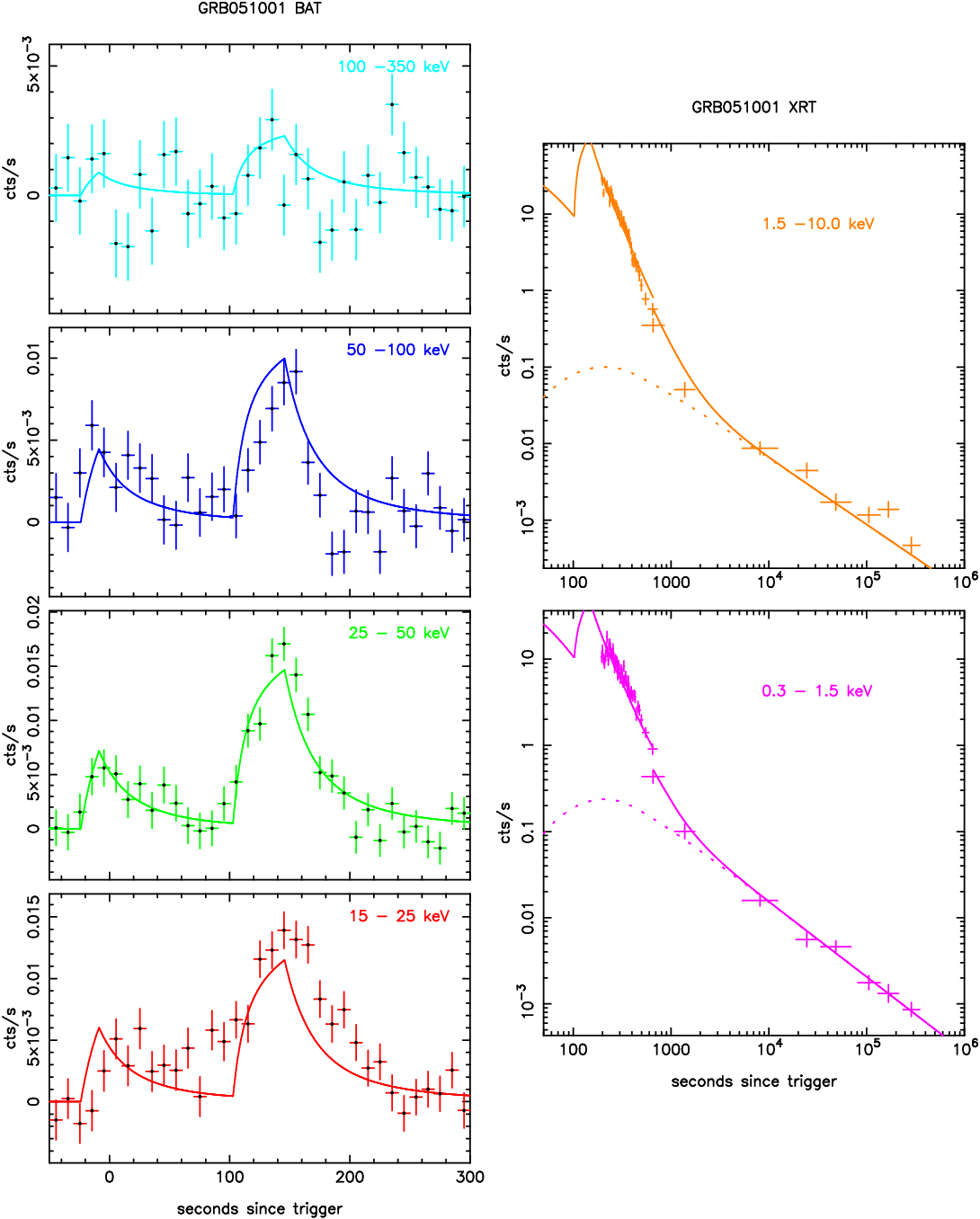}
\end{center}
\caption{GRB051001}
\label{fig4}
\end{figure}

This is an example of a short burst with an extended emission tail that is just visible
in the {\em Swift} BAT over a prolonged period such that $T_{90}=96$ seconds.
This extended emission is seen as a very broad weak rise in the BAT
count rate peaking at about 80 seconds and as a very early bright flare
in the XRT. This is
fitted by the 2nd pulse although it is
not possible to find a Band function spectrum which fits the BAT and XRT profiles exactly.
The fit shown in Fig. \ref{fig2} has the highest reduced
Chi-squared value of any of the fits. The model is
a little too low over most of the BAT 15-25 keV band
and doesn't quite match the shape of the 
tail of the RDP in the XRT 0.3-1.5 keV band.
For this essentially short burst the RDP observed is entirely due to the 
extended emission and not the initial short pulse.
Coincidentally, this bright, soft 2nd pulse gave rise to an expanding dust-scattered X-ray
halo around this GRB as reported by \citet{2006ApJ...639..323V}.
This burst also
has a late flare seen in the XRT which is included as the 3rd fitted pulse with
$T_{pk}=50000$ seconds and $T_{f}=43000\pm11000$ seconds. 

\subsection{GRB050814}

GRB050814 comprises a Fast Rise Exponential Decay (FRED) pulse followed
by a weaker 2nd pulse the decay of which
was tracked well by the XRT.
The fitted model can't match the short hard peak of the 
pulse (seen in the BAT 50-100 keV band) which decays rapidly combined with the
softer pulse which decays more slowly.
The pulse profile is not well fit in the BAT 15-25 keV band. The rapid decay of
the pulse is well fit by the predicted high latitude emission of the model in both XRT
energy bands except at the transition into the start of the afterglow
plateau. This is a good example where the data suggest a slower
decline before 100 seconds (in the BAT 15-25 keV band) but a faster
decline later, in the XRT bands.

\subsection{GRB051001}

A rather weak burst modelled by two pulses.
The second pulse is seen in both the BAT and the XRT
and the best fit is achieved using an $E_{f}=15\pm3$ keV,
which is in between  the XRT and BAT band passes.
As with other bursts the profile is not a good fit in the lowest
BAT band 15-25 keV.
There is also evidence for curvature of the decay of the 2nd pulse
in the XRT bands which is not matched by the model.

\subsection{GRB060211A}

This is a rather faint burst seen in the BAT which is reasonably well
represented by
two pulses except that the pulse profiles in the 50-100 keV band have rather 
flat tops with a sharp drop at the end which is not emulated by the model.
The fit for the 1st pulse in the 50-100 keV band is particularly
poor and it looks
as if the spectrum should be harder for this pulse. However, if the
spectral index is made smaller to accommodate this the decay of the
pulse is too shallow. i.e. it is not possible to satisfy the closure
between the spectral index and temporal index for this pulse.
The RDP seen in the XRT is very well matched by the model.

\subsection{GRB060814}

GRB060814 was a bright burst consisting of 3 well separated 
main pulses in the BAT seen with excellent
statistics. The best fit to the 2nd and 3rd pulse doesn't quite capture the narrow hard peak
seen in the BAT 100-350 keV band. The detailed shape of the
decay of the 2nd pulse is not well fitted and the residuals on this decay
contribute significantly to the Chi-squared value.
We tried varying the index $a$ but this did not improve the fit.
We also tried lifting the constraint $a+d=0$ and fitting $a$ and $d$ 
independently but this did not improve things either.
The 4th pulse is seen in both
the BAT and the XRT and it proved impossible to fit these data with a single
pulse. If we fit to the BAT only then $T_{f}=28$ seconds and the decay falls well 
below the XRT data for $t>150$ seconds.
Fitting both the BAT and XRT gives $T_{f}=113$ seconds but the peaks in the BAT
are missing and the fit is dominated by the XRT because there are many more data
points from the low energy instrument. It is not possible to find a spectrum
which can produce both the fast, narrow hard pulse and the wider soft decay.

\begin{figure}
\begin{center}
\includegraphics[height=15cm,angle=0]{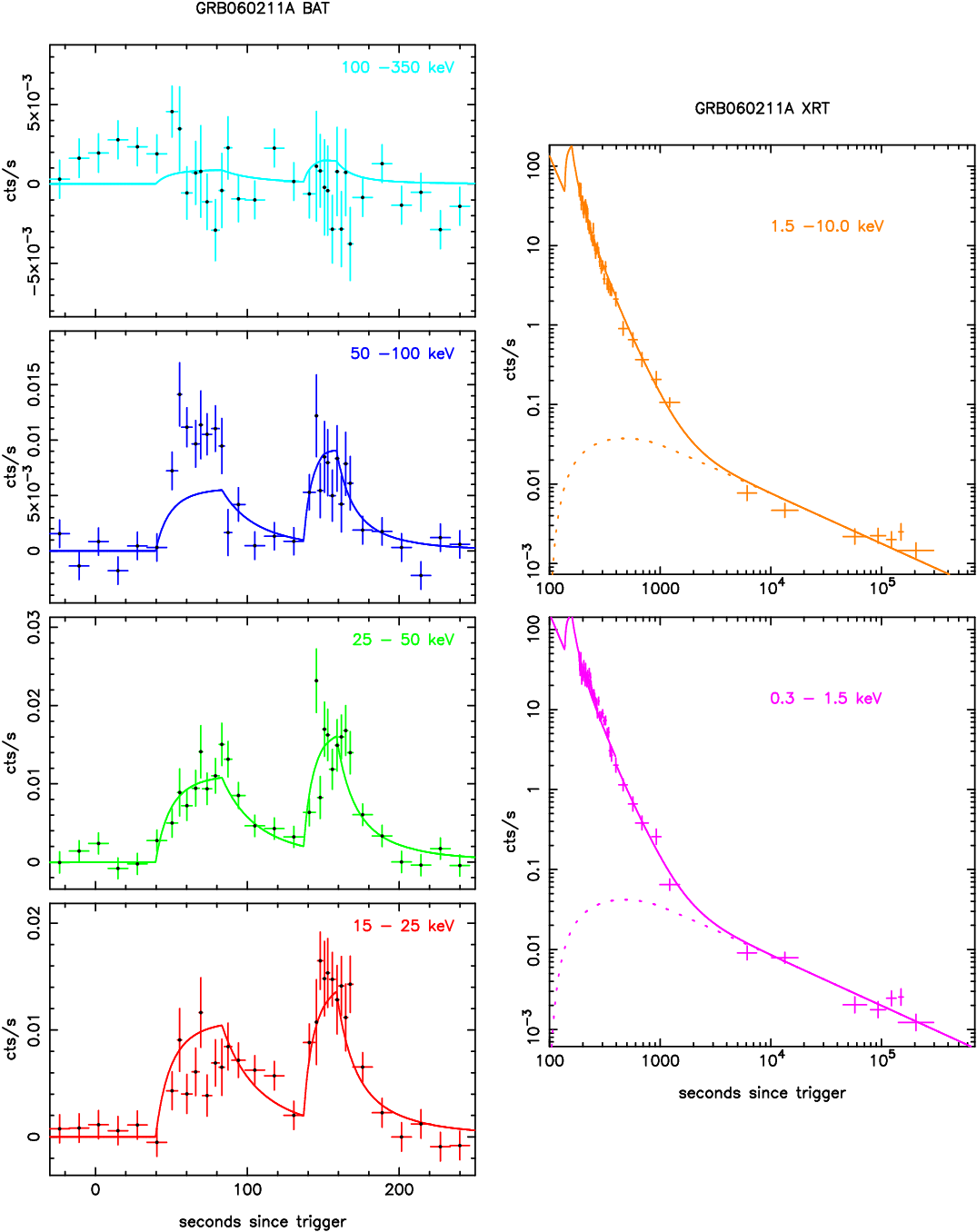}
\end{center}
\caption{GRB060211A}
\label{fig5}
\end{figure}

The fit shown in Fig. \ref{fig5} consists of a 4th hard pulse which fits
the peak seen in the
BAT and a slightly later peaking soft pulse which fits the decay seen in
the XRT.
If we calculate the ejection times $T_{ej}=T_{pk}-T_{f}$ for the 4th and 5th pulses
we get 92 seconds and -102 seconds respectively. Under the simple assumptions of the
model this is unphysical since the 5th pulse is launched before the 4th pulse (and indeed
before the 1st and 2nd pulses!) but the peak is seen after the 4th pulse.

\subsection{GRB061110A}

This burst comprises a single FRED pulse seen in the BAT followed by a soft pulse seen
only in the XRT. The pulse profiles and the decay into the afterglow are both well
fit by a simple two pulse model. There is a slight mismatch around the
transition from the rapid decay into the afterglow.

\subsection{GRB061121}

GRB061121 triggered on a well defined precursor which was
separated from the main body of the burst by about a minute. This gave
{\em Swift} time to slew onto source and so the XRT was able to capture
several pulses of the prompt emission alongside the BAT. The source
exhibited detailed spectral evolution as described by
\citet{2007ApJ...663.1125P}. The current model includes 6 pulses 3 of
which were seen in both the BAT and XRT. Because the statistics are
good the model is unable to reproduce all the features in the data
but the overall fit is reasonable. In particular,
the cusps of the peaks of pulses 3, 4 and 5 have rather large residuals
although they are difficult to see on the plot in \ref{fig8}.
The fifth and brightest pulse is
of particular note. The decay from the narrow peak is very rapid
but the spectrum is hard, $b_{1}=0.22\pm 0.04$ and close to a power law
over the full BAT and XRT range. It is impossible to
fit this decay using the standard model index $a=1.0$. The fit shown
in Figure \ref{fig8} uses $a=2.4$ (and hence $d=-2.4$).
A sixth pulse seen chiefly in the XRT interupts the decay but the
overall fit of the 
high latitude emission component from the pulses using this
large index value is good.

\subsection{GRB061222A}

A total of 8 pulses was used to model this bright burst and there are still weak features which are
not included in the model profile. The Chi-squared is dominated by the
residuals at the cusps of the pulses 5, 6 and 7 and the transition from
the rapid decay into the afterglow. 

\subsection{GRB070420}

The rather untidy BAT lightcurve of this burst has been fitted using
4 overlapping pulses. The rapid decay seen in the XRT 1.5-10.0 keV band is
not well modelled in the latter stages before it is obscured by
the plateau of the afterglow emission at $\approx 200$ seconds and
this makes a significant contribution to the Chi-squared but the
softer decay in the XRT 0.3-1.5 keV band is well matched.

\subsection{GRB070621}

This burst is rather different from the others presented here because
the lightcurve consists of a series of very similar overlapping pulses. The fit
shown in Figure \ref{fig11} comprises 6 pulses. For all these
the width $T_{rise}$ was fixed during the fitting.
The BAT data are very ragged
and clearly more complicated than suggested by the model. However,
once again the rapid decay captured by the XRT is well fit by the model.
The largest residuals occur just before the rise of the 1st pulse. There
is clearly some emission here which is not included in the model.

\subsection{GRB080229A}

This burst is a good example of a sequence of well defined pulses
including a precursor 25 seconds before the main burst. The model
is unable to fit the hard, narrow peak of pulses 1 (precusor), 2, 3, and 4.
The residuals at the cusps of pulses 3 and 4 a particularly large.

\subsection{GRB080805}

This burst is well represented by just two pulses, the trigger pulse
seen just in the BAT and a 2nd soft pulse seen in the XRT. The
1st pulse profile is well fit in the BAT 50-100 keV band but
the hard peak in the 100-350 keV band is not quite fit and there is
excess emission seen in the 15-25 keV band which peaks 20 seconds later.
We tried many different spectral profiles (other than the Band function) to
try and match this behaviour without success. The soft tail of this 1st pulse
seen at the start of the XRT observation in the 1.5-10.0 keV band is
also higher than predicted. Both pulses were better fit if $E_{f}$
was allowed to float and the resulting best fit values for this 
parameter are well constrained.

\begin{figure}
\begin{center}
\includegraphics[height=15cm,angle=0]{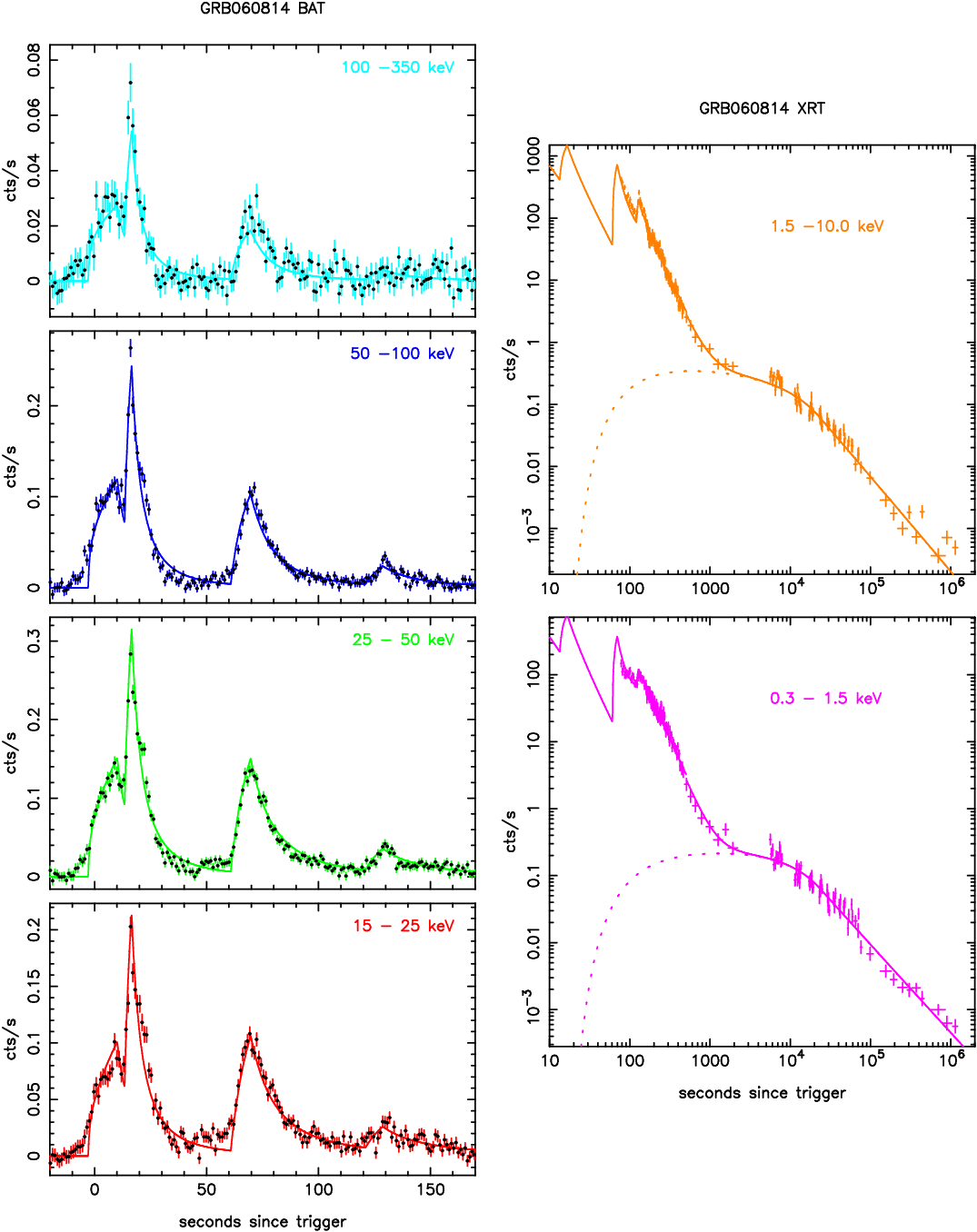}
\end{center}
\caption{GRB060814}
\label{fig6}
\end{figure}

\begin{figure}
\begin{center}
\includegraphics[height=15cm,angle=0]{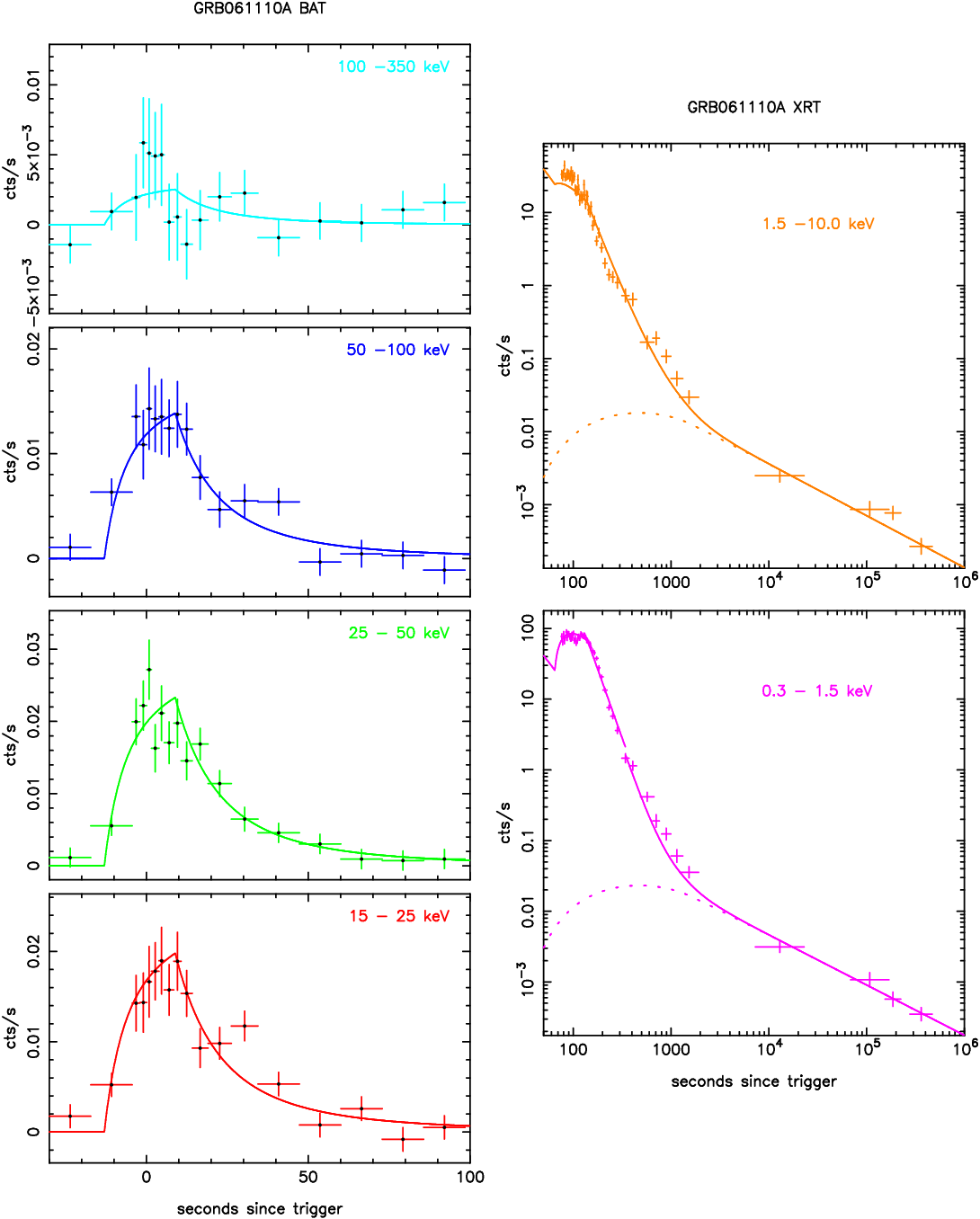}
\end{center}
\caption{GRB061110A}
\label{fig7}
\end{figure}

\begin{figure}
\begin{center}
\includegraphics[height=15cm,angle=0]{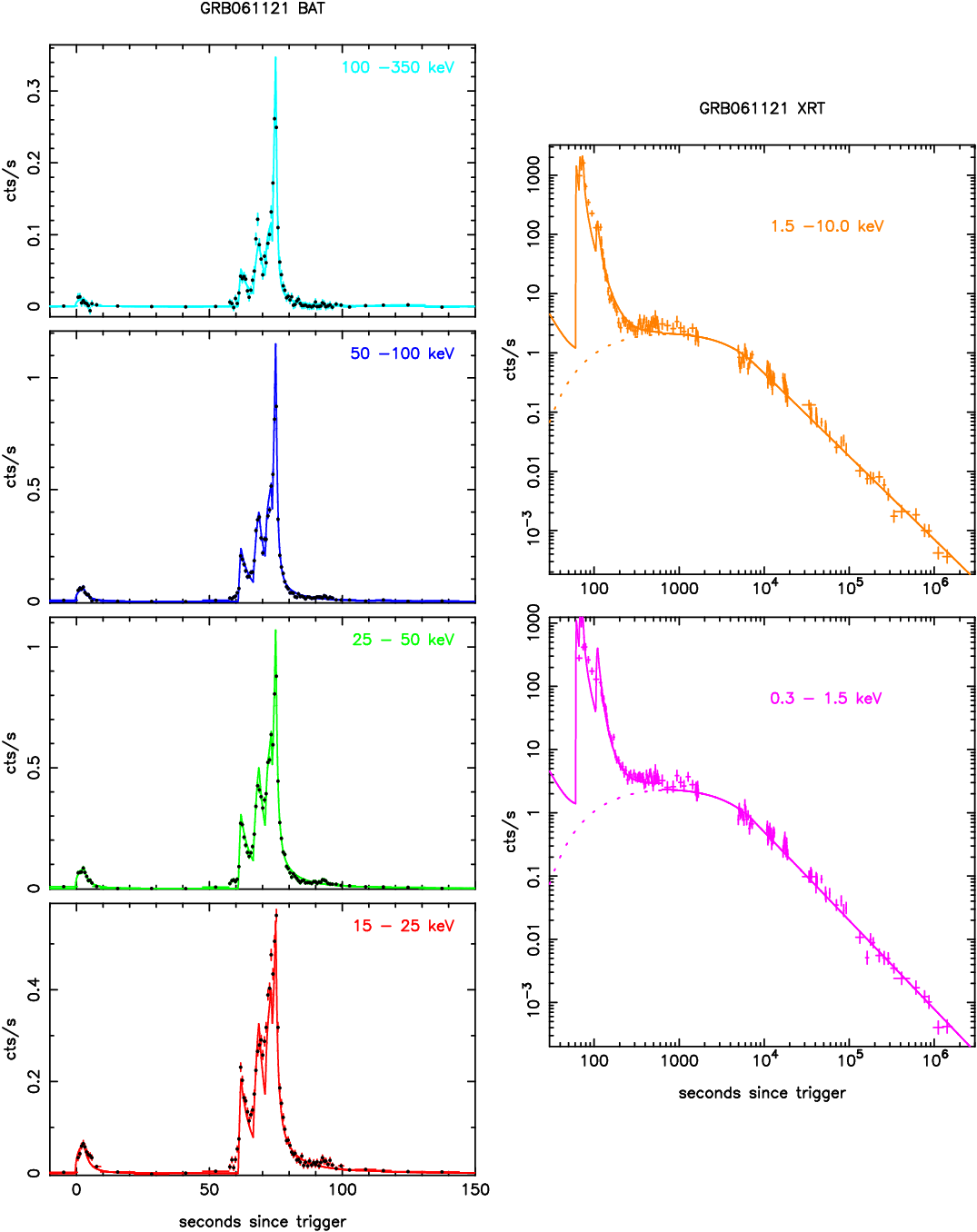}
\end{center}
\caption{GRB061121}
\label{fig8}
\end{figure}

\begin{figure}
\begin{center}
\includegraphics[height=15cm,angle=0]{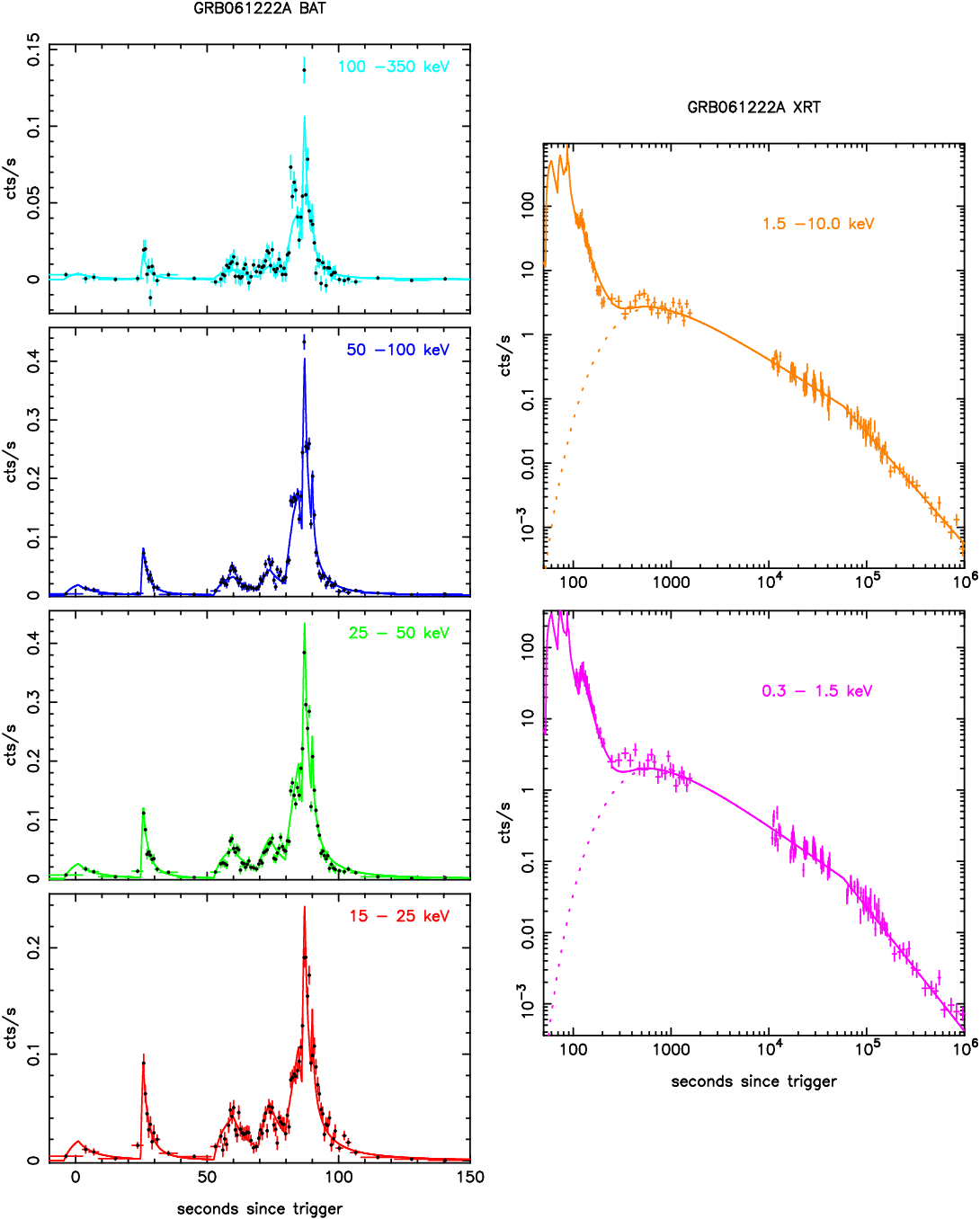}
\end{center}
\caption{GRB061222A}
\label{fig9}
\end{figure}

\begin{figure}
\begin{center}
\includegraphics[height=15cm,angle=0]{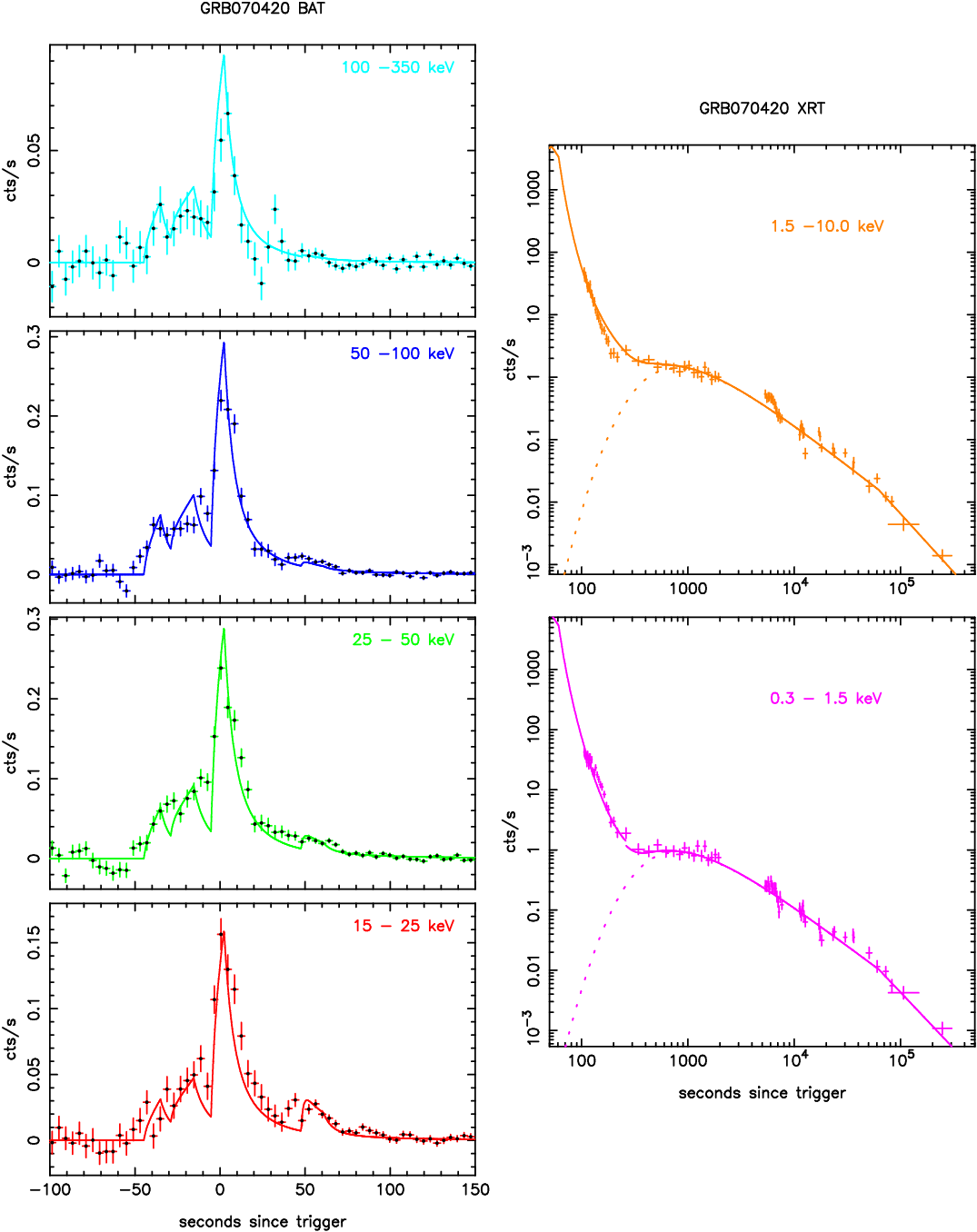}
\end{center}
\caption{GRB070420}
\label{fig10}
\end{figure}

\begin{figure}
\begin{center}
\includegraphics[height=15cm,angle=0]{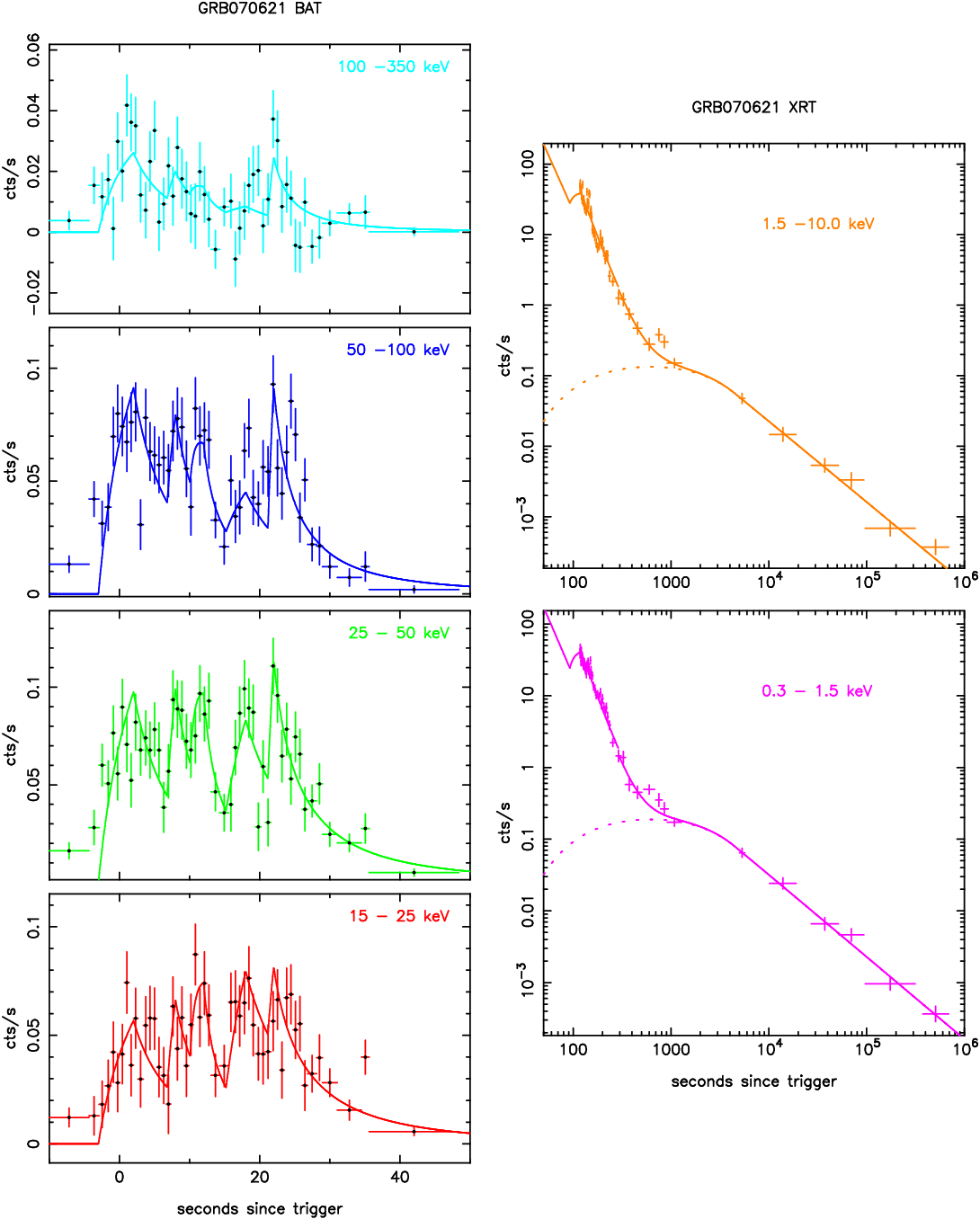}
\end{center}
\caption{GRB070621}
\label{fig11}
\end{figure}

\begin{figure}
\begin{center}
\includegraphics[height=15cm,angle=0]{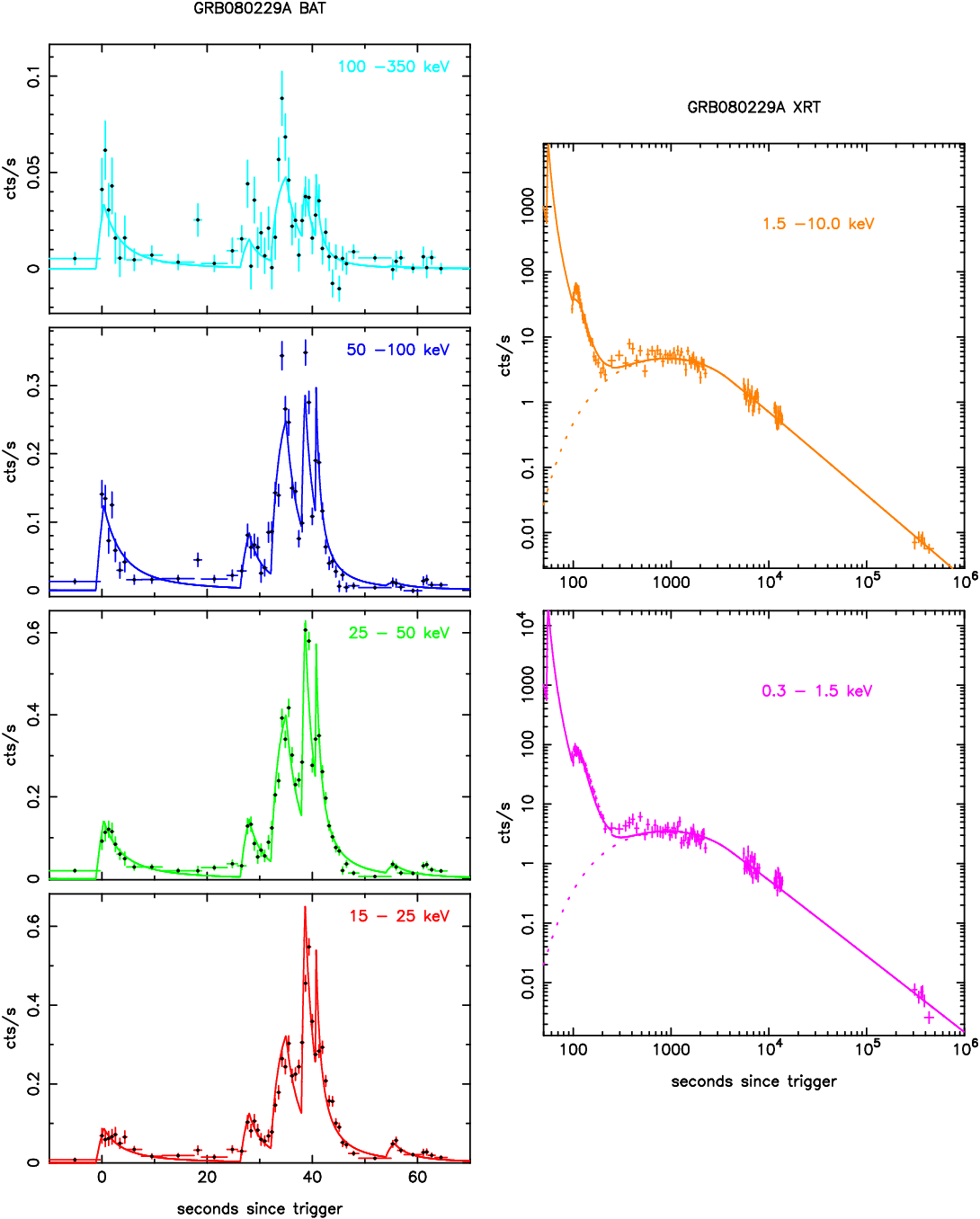}
\end{center}
\caption{GRB080229A}
\label{fig12}
\end{figure}

\begin{figure}
\begin{center}
\includegraphics[height=15cm,angle=0]{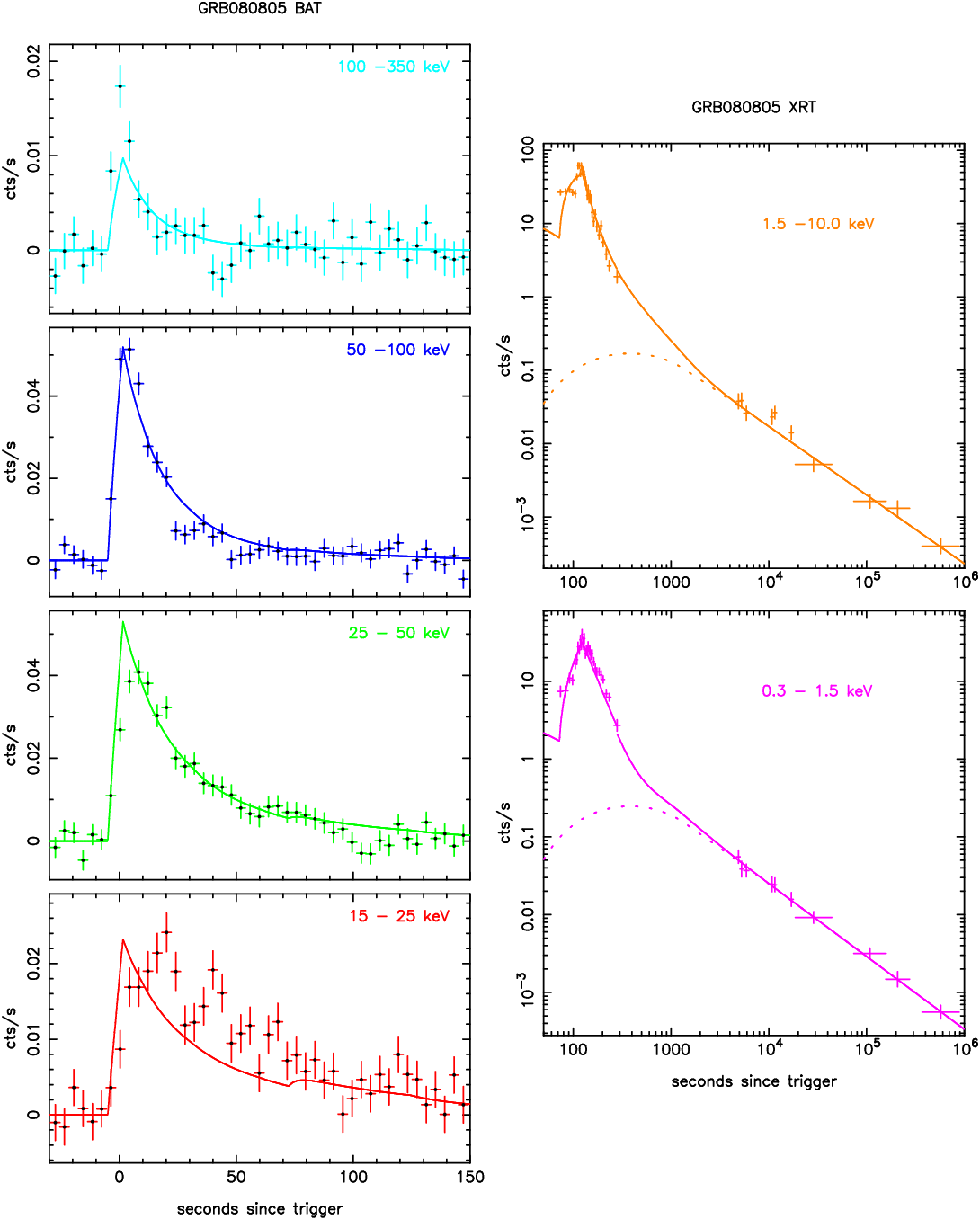}
\end{center}
\caption{GRB080805}
\label{fig13}
\end{figure}

\section{Discussion}

In all cases the fit to the prompt decay from the BAT into the XRT band
is reasonably good. In some cases like 050724, 051001 or 070420 there
is a residual curvature in the decay which is not exactly
matched by the model but in general the summation of the HLE
at late times from all the pulses provides a good fit to the RDP.

There is a generic problem with fitting hard, narrow peaks
in conjunction with a soft more slowly decaying tail.
The fourth pulse in GRB060814 is a prime example. Here it proved impossible
to fit the peak in the BAT 50-100 keV band, which produces the decay
seen in the XRT, using a single pulse.
A good fit can be obtained to just the BAT data using
a spectral index $b_{1}=-0.56$ but this is far too hard to produce the
soft tail seen in the XRT. Alternatively the XRT profile can be fitted
using $b_{1}=-1.09$ but this produces no visible flux in the BAT.
The fit shown in Figure \ref{fig6} includes a 5th pulse which fits the
broad soft tail seen in the XRT. There is a very significant difference between
the $T_{f}$ values of the 4th and 5th pulses in the final fit, $38\pm4$ and
$252\pm23$ seconds respectively. The required difference in the temporal
decay times in the hard BAT and soft XRT could not be satisfied using
a single spectral profile.
A similar sitution exists for the first pulse in GRB080805. The
profile in the upper three BAT bands is reasonable but this doesn't fit
the excess flux seen in the 15-25 keV BAT band or the start of the
XRT observation in the 1.5-10 keV band before 100 seconds when the
second pulse starts to dominate.

\begin{figure}
\begin{center}
\includegraphics[height=15cm,angle=0]{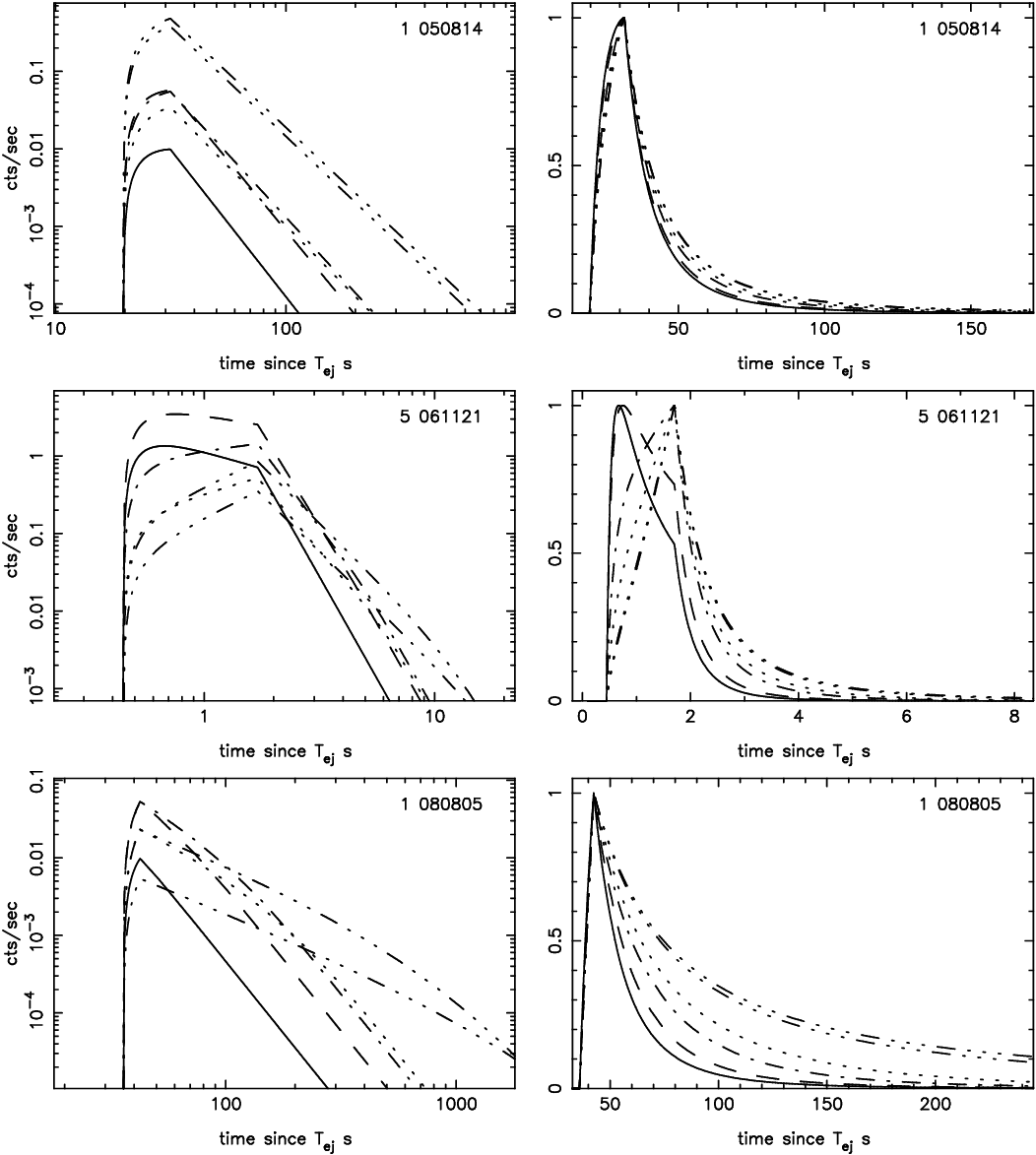}
\end{center}
\caption{Pulse profiles as a function of energy band. Top panels:
pulse 1 of 050814, a typical pulse.
Middle panels:
pulse 5 of 061121 which exhibits energy dependent lag.
Bottom panels: pulse 1 of 080805 which exhibits energy dependent pulse width.
Left-hand panels:
the count rate vs. time since $T_{ej}$ with logarithmic scales. The XRT
rates have been divided by a factor of 1000 to put them in the same range
as the BAT rates. Right-hand panels: The count rates have been scaled to
a fixed value of 1 at $T_{f}$ seconds after $T_{ej}$ and axes are
now linear. In all panels: BAT 100-350 keV solid, BAT 50-100 keV dashed,
BAT 25-50 keV dash-dot, BAT 15-25 keV dotted, XRT 1.5-10 keV and 
0.3-1.5 keV both dash-dot-dot-dot.}
\label{fig14}
\end{figure}

Despite these problems the model does
accommodate an element of energy-dependent lag \citep{2000ApJ...534..248N}
and narrowing of the pulses at high energies \citep{2001ApJ...552...57R} as
illustrated in Fig. \ref{fig14}. The top panels shows the 1st pulse
of 050814, a typical pulse. The characteristic energy at peak
is set to 1000 keV
so that the spectrum is very close to a simple power law over the BAT and XRT
energy range at the peak of the pulse.
The decay from the peak is a power law in each band, slightly
steeper for the upper energy bands closer to the characteristic energy.
The pulse
profile shown in linear scaling to the right is very similar in all energy
bands.
For the 5th pulse in 061121 shown in the middle panels of Fig. \ref{fig14}
the steep decay
requires $a=2.4$ (90\% range 2.1-2.8),
significantly higher than the standard value of $a=1$.
Because of curvature in the
spectrum (the exponential factor in the Band function) the high index also
has the effect of shifting the pulse peak to below $T_{f}$ for the harder
energy bands. The spectral lag introduced is clearly illustrated in 
Fig. \ref{fig14}.
The best fits for 8 of the pulses included $E_{f}$ as a fitted parameter.
For these pulses there is greater spectral curvature because
$E_{f}$ falls within the XRT-BAT energy range and consequently
the pulse profile
changes between the energy bands. The bottom panels of
Fig. \ref{fig14} show an example
of this; the 1st pulse in 080805. The rise of the pulse is almost
identical over all energy bands but the decay is steep at high energies
and shallow a low energies. The resulting pulse width is a strong function
of energy although the peak itself remains at the same position.
This is not a lag in the strictest sense of the term but
it would appear as a lag in
any analysis involving crosscorrelation since the centroid of the
peak clearly shifts to later times as the energy decreases.

The ability of the model to produce energy dependent lag and/or
pulse width is restricted by the requirement for closure between
the temporal decay index and spectral index. If a pulse is bright
in the hard energy band the spectral index tends to be close to zero
but this in turn tends to produce a slow initial decay of the hard pulse.
Conversely, if the pulse is soft then the spectral index is large and
negative and we expect a fast initial decay in the soft band.
It may be possible to fit the lag/narrowing seen in pulses like the
1st pulse of 080805 or the hard peak of the 2nd pulse of 060814 using
two pulse components rather than just one. However such component pulses
would overlap to a large degree and the parameters of the pulses should
be coupled in the fitting procedure to prevent the result becoming
unphysical. This was not attempted in the current work.

Our model is rather simplistic and doesn't include specific physical
details associated with, for example, the internal shocks.
When two shells collide there will be two emission regions, corresponding to
the shocked portions of each shell, and they can have
the same onset time, $T_{ej} + T_{0}$, the same $R_{0}$ and
$\Gamma$ ($R\ge R_{0}$) but different emission conditions,
$E_{peak}$ or $\Delta R$, and different peak times, $T_{ej} + T_{f}$.
We assume the simplest
possible history for each shell, that it has coasted at the same $\Gamma$
since it was ejected, and retains it also when emitting, while in practice
multiple collisions are possible and the
value of $\Gamma$ during the emission can be different than before. It can
be increased or decreased due to the collision with the second shell
that triggers the emission or some deceleration of the front shell
due to the sweeping-up of the external medium may also occur.
The emission may not be caused by internal shocks and may not be
synchrotron so the values of indices $a=1$ and $d=-1$ might
be inappropriate.
All these factors will clearly modify the pulse profile and spectrum
and the way pulses overlap.

\begin{figure}
\begin{center}
\includegraphics[height=15cm,angle=-90]{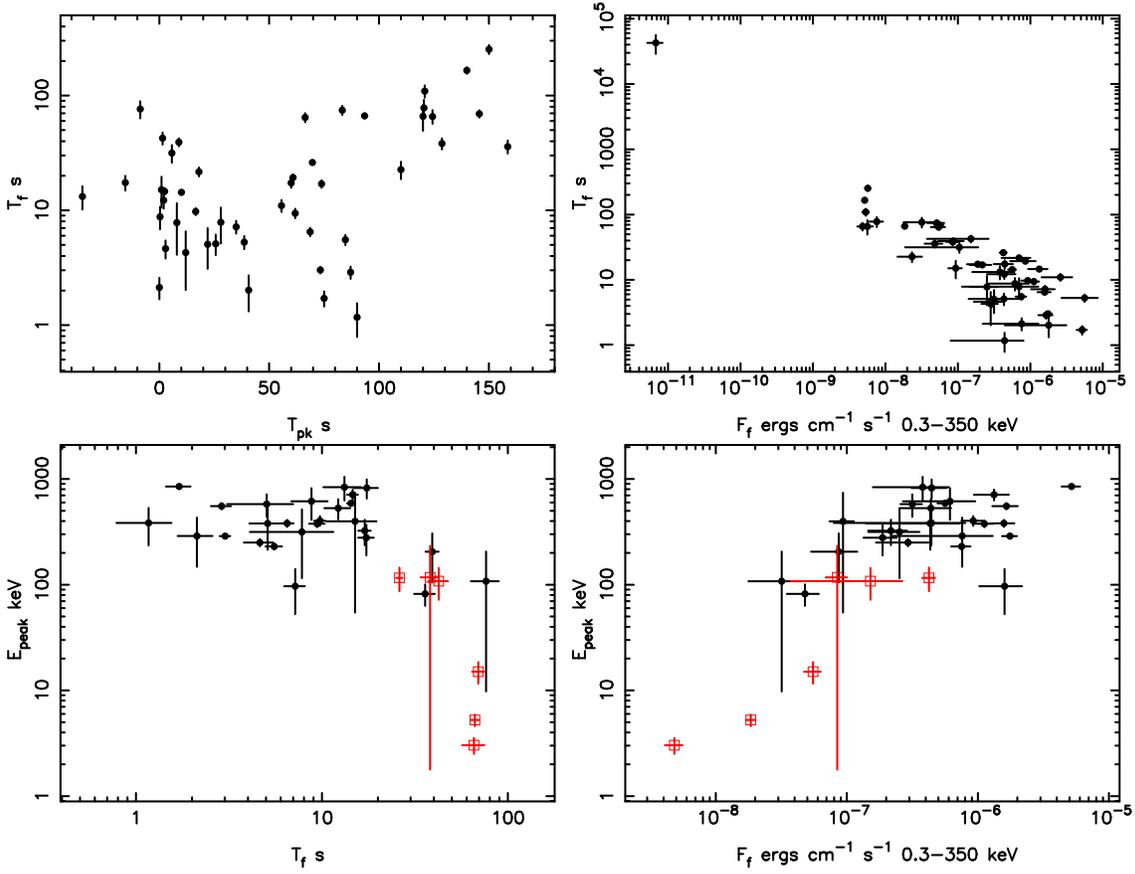}
\end{center}
\caption{The fitted parameter distributions. Top left-hand panel:
The characteristic
pulse time, $T_{f}$, vs. the position of the peak with respect to
trigger time, $T_{pk}$.
The late flare in 050724 has been omitted because $T_{pk}$ can be negative
and therefore cannot be plotted on a logarithmic scale. Top right-hand panel:
$T_{f}$ vs. the peak flux, $F_{f}$.
Bottom left-hand panel: The
peak energy $E_{peak}$ vs. $T_{f}$.
Bottom right-hand panel: The
peak energy $E_{peak}$ vs. $F_{f}$. The fits for which 
an error range in $E_{f}$ was calculated are shown as squares ( red).}
\label{fig15}
\end{figure}

There are three parameters which were allowed to float in the fitting
for all the pulses; the peak flux $F_{f}$, the characteristic time $T_{f}$ and
the spectral index $b_{1}$. We have used $b_{1}$ in conjunction with
$E_{f}$ to calculate the peak energy $E_{peak}$ which, as introduced in
\citet{1993ApJ...413..281B}, is a measure of the spectral hardness.
The distribution of these parameter values
are shown in Figure \ref{fig15}. The top left-hand panel shows $T_{f}$
vs. the peak time $T_{pk}$. Somewhat surprisingly there is no
strong correlation between the position of the peak with respect to
the trigger time
and length of the pulse although there is a consistent 
trend for the later peaks,
$T_{pk}>50$ seconds. So within the prompt phase the time scale of the pulses as
measured by $T_{f}$ is not dictated by the time since the start of the burst.
The sequence of $T_{f}$ values for 061121 is a good example of this.
The top right-hand panel shows the correlation between
$T_{f}$ and the peak flux $F_{f}$. Here there is an
obvious trend which includes the late, weak flare seen by the XRT in 050724.
Similar correlations were discussed by \citet{2000ApJS..131....1L} and
\citet{2002A&A...385..377Q} although they were considering the pulse
width rather than $T_{f}$. We demonstrate later in this section that
$T_{f}$ is in fact closely correlated with pulse width.
The first survey of X-ray flares
seen in GRBs, \citet{2007ApJ...671.1903C},
reports a correlation between
the flare flux and time of flare since the trigger. Again, this is clearly
related to the $F_{f}-T_{f}$ correlation although we have shown that $T_{f}$
is not well correlated with the time since trigger for early pulses.
The bottom panels show the
peak energy  $E_{peak}$, plotted against $T_{f}$ (left-hand panel)
and $F_{f}$ (right-hand panel). The points for which an error
range in $E_{f}$ was calculated are marked in red.
For $T_{f}>10$ the peak energy clearly drops with increasing $T_{f}$ and
$E_{peak}$ is also correlated with the peak flux, $F_{f}$.

\begin{figure}
\begin{center}
\includegraphics[height=14cm,angle=-90]{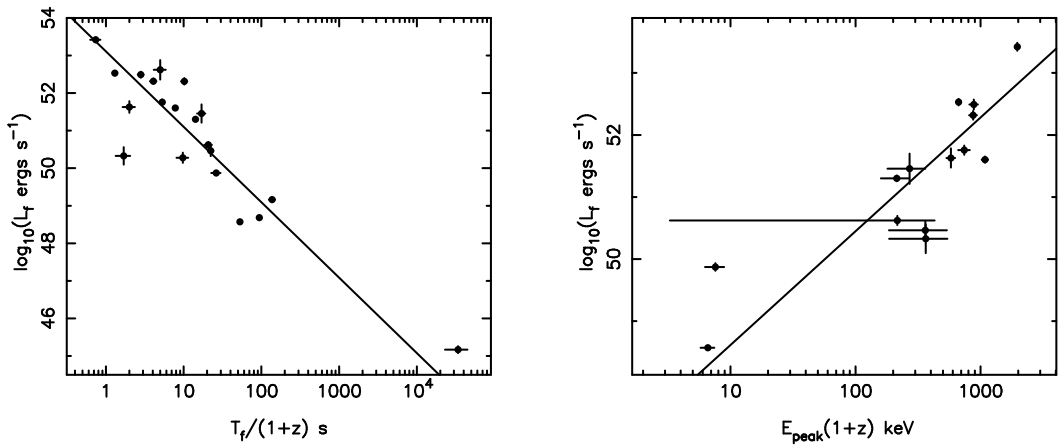}
\end{center}
\caption{Left-hand panel: The luminosity at the peak of the pulse
$L_{f}$ ergs s$^{-1}$ vs. 
the characteristic pulse time in the source frame, $T_{f}/(1+z)$.
Right-hand panel: The luminosity at the peak of the pulse
$L_{f}$ ergs s$^{-1}$ vs.
the peak energy of the spectrum in the source frame,
$E_{peak}(1+z)$.}
\label{fig16}
\end{figure}

Table \ref{tab2} also includes an estimate of the peak
luminosity of the pulses for those bursts with a measured redshift. Because
the peak flux, $F_{f}$, is correlated with $T_{f}$, as shown in Figure
\ref{fig15},
the luminosity, $L_{f}$, is similarly correlated with the characteristic
pulse time in the source frame, $T_{f}/(1+z)$, as shown in Figure \ref{fig16}.
The best fit correlation line plotted is
$L_{f}=1.3\times10^{53}(T_{f}/(1+z))^{-2.0\pm0.2}$ ergs s$^{-1}$.
The most luminous pulse is the 5th pulse in 061121 with
$T_{f}/(1+z)=0.7$
seconds and $L_{f}=2.6\times10^{53}$ ergs s$^{-1}$
while the least luminous prompt pulses (ignoring the
late flare in 050724) were soft and only seen in the XRT, for example
the 2nd pulse of 050724 with $T_{f}/(1+z)=53$ seconds and
$L_{f}=3.7\times10^{48}$ ergs s$^{-1}$.
A very similar correlation between X-ray flare luminosity and time of
flare since the GRB was presented by 
\citet{2008MNRAS.388L..15L}.  They were considering the time since
trigger of the peak of the flare, referred to as $T_{pk}$ here,
but for late flares seen in the XRT this is very similar to $T_{f}$.
A correlation between $T_{f}$ and $T_{pk}$
for the later prompt pulses, $T_{pk}>100$ s, is evident 
in the top left-hand panel of Figure \ref{fig15}.
Assuming that $T_{pk}$ is a good proxy for $T_{f}$ for the flares
analysed by \citet{2008MNRAS.388L..15L}
the values derived here extend the
relation down to $T_{f}/(1+z)=1$ second. From this perspective
the prompt pulses and late X-ray flares may form a continuum consistent with
a common origin associated directly with activity in the central engine.
We note that the powerlaw index derived by
\citet{2008MNRAS.388L..15L}, $\alpha=-1.5\pm0.16$, is a little shallower
than the value of $-2.0\pm0.19$ here although this may be due to the small
number of pulses/flares included in the analysis. 
For early prompt
pulses there is no correlation between $T_{pk}$ and $T_{f}$ (Figure \ref{fig15}
top left-hand panel) so the extended correlation only makes
sense if we consider the time since ejection of the pulses not
the time since the start of the burst. If this is the case the
arguments that the correlation is associated with accretion,
$L\propto \dot{m} \propto t^{-1.2}$ or alternatively with 
the spin-down of a magnetar, $L\propto t^{-2}$, are not appropriate.
A more detailed analysis incorporating a far larger number of prompt pulses
and flares is required to see if they really do obey a simple powerlaw
correlation over the entire range $1<T_{f}<10^{4}$ seconds or whether the prompt
pulses and late flares form two distinct populations indicating that
they arise from different mechanisms.

A significant correlation of $L_{f}$ with the peak energy of the spectrum
in the source frame,
$E_{peak}(1+z)$, is also shown in Figure \ref{fig16}. The
best fit correlation plotted is
$L_{f}=6.1\times10^{46} (E_{peak}(1+z) keV)^{1.83\pm0.16}$ ergs s$^{-1}$.
This correlation is similar to the luminosity-peak energy relation
reported by \citet{2004ApJ...609..935Y} but here we stress that
we are considering individual pulses and not the average behaviour of bursts.
The correlation is between the luminosity at a particular
time ($T_{f}$ seconds after ejection when the emission suddenly turns off)
and the peak energy at exactly the same time indicating a strong
physical link between the these quantities.
A spectral-luminosity relation within individual {\em Fermi} GRBs has also
been reported by \citet{2009arXiv0908.2807G}. The correlation they find
is consistent with the values reported here and with the peak-energy
luminosity correlation for long bursts
discussed by \citet{2008MNRAS.391..639N}.
\citet{2009ApJ...704.1405K} provide spectral fits with estimates
of $E_{peak}$ for time intervals within
individual bursts (see Table \ref{tab3} and the discussion above) but
they concern themselves with the $E_{peak}$-$E_{iso}$ relation rather
than correlation with the peak luminosity.

The rise time of the pulse, $T_{rise}$, was also included in the fitting
although in many cases it was fixed because the data were not good
enough to determine a useful error range.
\begin{figure}
\begin{center}
\includegraphics[height=15cm,angle=-90]{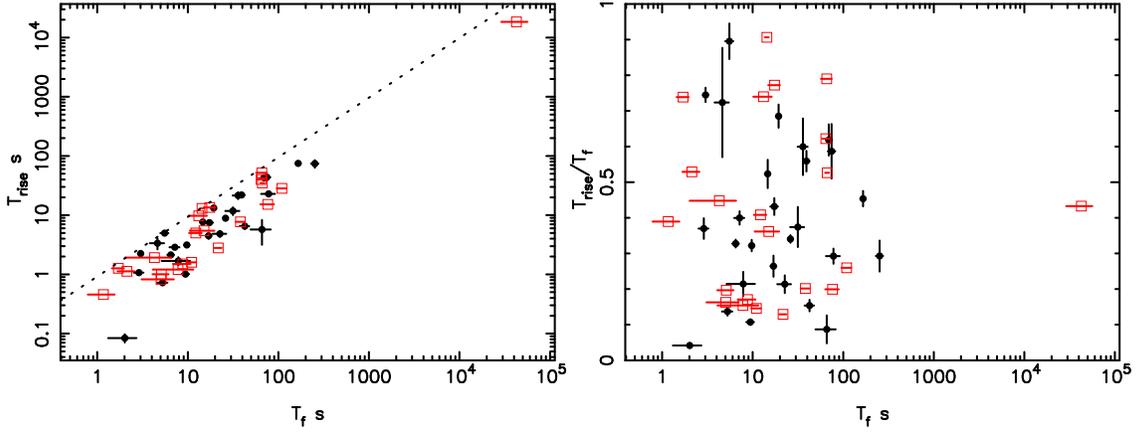}
\end{center}
\caption{The rise time parameter $T_{rise}$. 
Left-hand panel: $T_{rise}$ vs. $T_{f}$. The dotted line indicates
equality.
Right-hand panel: The ratio $T_{rise}/T_{f}$ vs. $T_{f}$.
The points with no error bars for $T_{rise}$
(i.e. fixed in the fitting) are plotted as squares (red).}
\label{fig17}
\end{figure}
The left panel of Fig. \ref{fig17} shows a strong correlation between the
rise time and the characteristic time $T_{f}$. As expected
$T_{rise}$ is always
less than $T_{f}$ in accordance with the definitions illustrated in
Fig. \ref{fig1}. This result was not included as a constraint in the
fitting but arises from the shape of the pulses fitted. Because of this
strong correlation the distribution of $T_{rise}$ with respect to the
other parameters are very similar to those of $T_{f}$ shown
in Fig. \ref{fig15}.
The right-hand panel of Fig. \ref{fig17}
shows the distribution of the ratio $T_{rise}/T_{f}$
with respect to $T_{f}$.
This ratio is closely related to the distance the shell moves during the
emission relative to the radius of the shell at the start of the emission
\citep{2009MNRAS.399.1328G},
$T_{rise}/T_{f}=\Delta R/(R_{0}+\Delta R)$. There is no correlation
between this ratio and the other parameters.

\begin{figure}
\begin{center}
\includegraphics[height=15cm,angle=-90]{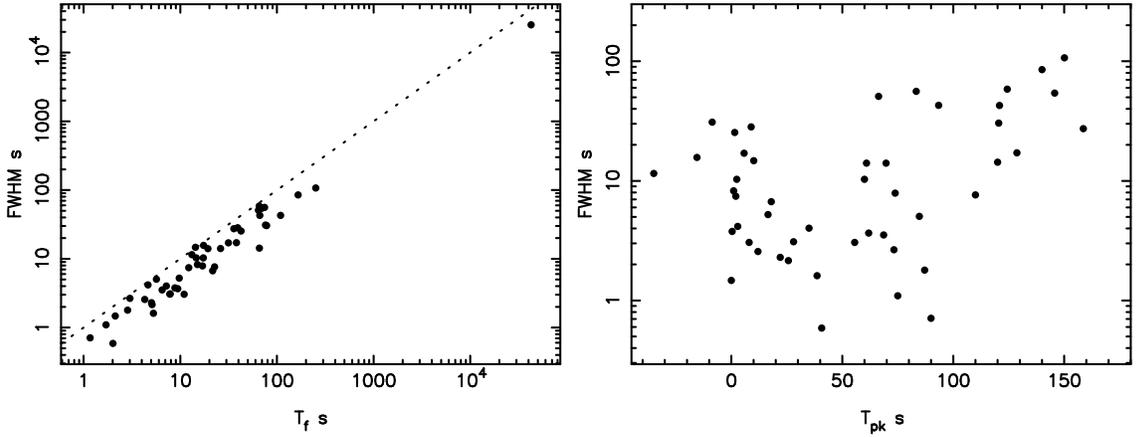}
\end{center}
\caption{The Full Width at Half Maximum of the pulses in the BAT
energy band 15-25 keV. 
Left-hand panel: The FWHM vs. the characteristic time $T_{f}$.
The dotted line indicates equality.
Right-hand panel: The FWHM vs. the time since trigger of the peak, $T_{pk}$.}
\label{fig18}
\end{figure}

The time from the ejection of the shell to its peak 
is physically the time when the emission from the shell
switches off at radius $R_{f}$ and thus determines the decay time
of the peak, which in turn usually dominates the width of the pulse.
So while $T_{f}$ cannot be measured directly because the ejection
of the shell is not observed the width of the pulse largely dictates
the value of $T_{f}$ in the fitting and we expect this time to be
correlated with the Full Width at Half Maximum (FWHM) of the pulses
which is directly measureable.
The FWHM of the pulses is a complicated function of $T_{f}$, $T_{rise}$,
the spectral parameters $b_{1}$, $E_{c}$ and $b_{1}-b_{2}$ and the
observed spectral band so we have estimated it by numerical interpolation
for the lowest energy band of the BAT. The left-hand panel of
Fig. \ref{fig18} shows the FWHM plotted against $T_{f}$ clearly
showing the expected correlation. Note that the FWHM is always less or equal to
$T_{f}$. The right-hand panel shows the FWHM vs. the time of peak since
the trigger, $T_{pk}$. In this case there is no correlation because
the pulse width or $T_{f}$ are not related to the time since the start of
the burst. Because of the tight correspondence between pulse width
and $T_{f}$ any correlations involving $T_{f}$ presented here 
also hold if we substitute the observed FWHM for the fitted value of
$T_{f}$. Similarly we expect the correlations
and analysis involving the FWHM of flares given elsewhere, for example in
\citet{2007ApJ...671.1903C}, could be interpreted in terms of $T_{f}$.

\section{Conclusions}
To summarise, we have successfully fitted the spectrally resolved light
curves of 12 GRBs using an analytical expression for pulse
profiles derived from the internal shock model incorporating the
spectrum and decay indices expected from synchrotron emission in the
fast cooling regime and a
decay tail arising from HLE. A total of 49 pulses describe all the major
features through the prompt emission into the RDP and, in one case,
a late X-ray flare. In all cases the fit to the RDP is good and there
is no doubt that the HLE can account for this phase of the emission
from GRBs. In many cases the energy dependent pulse profile
is as predicted by the model showing elements of spectral lag
and spectral pulse broadening and indicating that the spectral evolution
incorporated in the model is in agreement with observations.
There are aspects of the data which are not in accordance with the simple
model. One short, hard pulse in 061121 required index $a=2.4$, much
higher than the value expected for synchrotron emission, $a=1$. Several
pulses have a hard peak which cannot be adequately fitted. There is one
pulse, the 4th in 060814, which has a narrow hard peak and broad soft
tail which could not be represented by one pulse but had to be split
into two components. In the present procedure the pulses were fitted
independently with no constraints between the parameters of
individual pulses. Future modelling should include such constraints so
that the overlap of the pulses remains physically possible.
We find that the luminosity of pulses is anti-correlated with the
time since ejection for the pulses such that bright pulses arise
when the time since ejection is small while the dim pulses correspond to
long times since ejection. The individual pulse luminosity is also correlated
with the peak energy of the pulse spectrum consistent with the known correlation 
of peak luminosity for the entire burst with the peak energy of the $T_{90}$ 
spectrum.

\section*{Acknowledgments}
RW and PTO gratefully acknowledge STFC funding for {\em Swift} at the
University of Leicester. JG gratefully acknowledges a Royal Society
Wolfson Research Merit Award.

\begin{bibliography}{pulse_fitting_r1}
\bibliographystyle{mn2e}
\end{bibliography}

\label{lastpage}
\end{document}